\documentclass[twocolumn,pra]{revtex4}
\usepackage{amssymb}
\usepackage{amsfonts}
\usepackage{amsmath}
\usepackage{graphicx}

\begin{document}

\title{Quantum-classical hybrid dynamics: coupling mechanisms and diffusive
approximation}
\date{\today }
\author{Adri\'{a}n A. Budini}
\affiliation{Consejo Nacional de Investigaciones Cient\'{\i}ficas y T\'{e}cnicas
(CONICET), Centro At\'{o}mico Bariloche, Avenida E. Bustillo Km 9.5, (8400)
Bariloche, Argentina, and Universidad Tecnol\'{o}gica Nacional (UTN-FRBA),
Fanny Newbery 111, (8400) Bariloche, Argentina}

\begin{abstract}
In this paper we demonstrate that any Markovian master equation defining a
completely positive evolution for a quantum-classical hybrid state can
always be written in terms of four basic coupling mechanisms. Each of them
is characterized by a different \textquotedblleft
backaction\textquotedblright\ on each subsystem. On this basis, for each
case, we find the conditions under which a diffusive limit is approached,
that is, the time evolution can be approximated in terms of the first and
second derivatives of the hybrid state with respect to a classical
coordinate. In this limit, the restricted class of evolutions that guaranty
the positivity of the hybrid state at all times (quantum Fokker-Planck
master equations) emerges when the coupling mechanisms lead to infinitesimal
(non-finite) changes in both the quantum and classical subsystems. A broader
class of diffusive evolutions is obtained when positivity is only granted
after a transient time or alternatively is granted after imposing an initial
finite width on the state of the classical subsystem. A set of
representative examples support these results.
\end{abstract}

\maketitle

\section{Introduction}

Classical non-equilibrium dynamics are characterized by phenomena such as
dissipation and diffusion. These emergent dynamical effects admit a simple
description when approaching a Markovian limit. In fact, in this regime the
main descriptive theoretical tools are classical master equations,
Fokker-Planck equations, as well as their underlying description in terms of
stochastic Langevin equations~\cite{kampen,gardiner,Risken}.\ In a quantum
regime, since the underlying dynamics takes place in a Hilbert space,
non-equilibrium dynamics are characterized by dissipation, but also by
decoherence. In a Markovian regime, the theory of open quantum systems~\cite
{petruccione} allows these effects to be described in terms of Lindblad
(Lindblad-Kossakovski-Sudarshan-Gorini) dynamics, that is, quantum master
equations~\cite{petruccione,alickibook}. The stochastic representation of
these evolutions intrinsically involves which measurement process is
performed on the system of interest~\cite
{carmichaelbook,wisemanBook,barchielliBook}.

\textit{Quantum-classical hybrid systems} are bipartite arrangements that
lie on the frontier between the above ones, that is, the nature of one
subsystem is classical~\cite{kampen,gardiner,Risken} while the other is
quantum~\cite
{petruccione,alickibook,carmichaelbook,wisemanBook,barchielliBook}. Their
description involves a hybrid state that lies on a Hilbert space which in
turn depends parametrically on the state of the classical (incoherent)
system.\ Different physical situations motivated the study of hybrid
dynamics such as for example measurement theory~\cite{measurement},
Bloch-Boltzmann equations~\cite{alickipaper}, and non-Markovian master
equations~\cite{vega} induced by complex structured reservoirs~\cite
{esposito,random,breuerPreSpin}. In the latter context, information encoded
in the classical subsystem acts as a source of memory effects for the
partial quantum dynamics.

For Hilbert spaces of arbitrary dimension and \textit{discrete} (even
infinite countable) classical degrees of freedom the most general
time-evolution of the hybrid state was determine in Ref.~\cite{rate} and
also~\cite{breuerRate}. Due to their quantum-classical structure the
corresponding evolutions were termed as \textquotedblleft Lindblad rate
equations\textquotedblright ~\cite{rate} and alternatively as
\textquotedblleft generalized Lindblad master equations\textquotedblright ~
\cite{breuerRate}. A similar characterization was subsequently presented in
Ref.~\cite{sudarshan}. Specific applications were found in the context of
spin baths~\cite{spinbaths} and single-molecule spectroscopy~\cite
{barkaiChem}, that is, the study of (quantum) fluorescent systems coupled to
(classically) structured environments~\cite{sms,smsJumps} in which the
scattered electromagnetic field is usually observed by photon-detection
measurements~ \cite{smsJumps,rapidPoisson,light}. Stochastic representations
of the hybrid state were also studied from different perspectives~\cite
{smsJumps,rapidPoisson,light,barchieliJump,petruccioneJump}. The quantum
regression theorem~\cite{carmichaelbook,QRT}, quantum state smoothing~\cite
{smoothWise} to improve the efficiency of a photon-detector~\cite{smoothRate}
as well as the thermodynamics induced by finite baths~\cite{sanpera,pekola}
were also physical applications of hybrid dynamics studied in recent
literature.

The previous advancements in the study of quantum-classical hybrid dynamics
are mainly based on considering classical subsystems defined by a countable
set of possible states, while the quantum subsystem is arbitrary. A related
but different research line\ is defined by the complementary case, that is,
when the classical degrees of freedom can be labeled by a \textit{continuous
real coordinate}. In this situation, it is possible to endow the evolution
of the classical (incoherent) subsystem with its own Hamiltonian dynamics
(symplectic structure). These elements arise, for example, in the study of
physicochemical processes~\cite{kapral}. Morevoer, this type of hybrid
dynamics has been obtained by considering continuous-in-time measurement
processes~\cite{halliwel,strunz,diosi}. Furthermore, they have been proposed
as a possible model to describe the interaction between quantum matter and
classical gravitational fields~\cite{karol,penrose,LajosConfer}. Phenomena
such as a gravity induced decoherence emerge in this situation~ \cite
{karol,penrose,LajosConfer,bassi,tilloyDiosi,oppenGravity1,oppenGravity2}.
Notably, the most general time-evolution for an hybrid state with a
continuous classical coordinate was recently established in Refs.~\cite
{oppen1,oppen2} and lightened in~\cite{lastDiosi}. Interestingly, with a
completely different motivation, the resulting \textquotedblleft quantum
Fokker-Planck master equation\textquotedblright\ was also obtained in Ref.~ 
\cite{QFP} on the basis of a measurement-feedback protocol.

Despite the above advances, some features or questions about hybrid dynamics
remain open. First, we ask what is the minimal set of possible
quantum-classical coupling mechanisms that allow any hybrid evolution to be
written in terms of them. This question has not yet been clearly
characterized. Second, we ask under which conditions the evolution of the
hybrid state can be approximated by a diffusive one. The main goal of this
paper is to answer these questions.

Similarly to Refs.~\cite{rate,sudarshan}, the present approach is based on
embedding the hybrid state in a bipartite Hilbert space. Furthermore, to
address the above issues, the bipartite state is assumed to obey a local-in
time (Markovian) completely positive Lindblad dynamics~\cite
{petruccione,alickibook}. After imposing that one of the subsystems is
incoherent at all times in a given fixed basis, the bipartite Lindblad
structure allow us to characterize all possible quantum-classical coupling
mechanisms. The \textit{backaction}, that is, the impact of each subsystem
on the other, is also studied in detail.

The developed approach leads to hybrid dynamics where the classical
subsystem is characterized by a set of discrete states. On this basis,
similarly to classical systems where diffusive (Fokker-Planck) equations are
derived from (discrete) master equations~\cite{kampen}, we study under what
conditions a diffusive limit is approached. Quantum Fokker-Planck master
equations that preserve the positivity of the hybrid state at all times~\cite
{oppen1,oppen2,lastDiosi,QFP} emerge in the particular case in which the
coupling mechanisms lead to \textquotedblleft
infinitesimal\textquotedblright\ (smallest scale of the problem)
transformations in both subsystems. However, we also show that a broader
family of evolutions can be consistently taken into account after
disregarding an initial transient time or by restricting the wide of the
classical subsystem initial state. A set of representative examples supports
the main conclusions.

The paper is outlined as follows. In Sec. II we derive the set of all
possible quantum-classical coupling mechanisms that define any local-in-time
hybrid evolution. In Sec. III we analyze under which conditions a diffusive
limit emerges. In Sec. IV we study a set of specific examples. In Sec. V we
provide the Conclusions.

\section{Quantum-classical coupling mechanisms}

We consider a bipartite arrangement consisting of two parts, the quantum
subsystem ($s$) and extra degrees of freedom that in a second step will be
associated to the classical subsystem\ ($c$). In correspondence, the total
Hilbert $\mathcal{H}_{sc}$ is the product of each subsystem Hilbert space $
\mathcal{H}_{sc}=\mathcal{H}_{s}\otimes \mathcal{H}_{c}.$ By assumption, the
density matrix $\Xi _{t}$ of the bipartite arrangement obeys a Lindblad
equation~\cite{petruccione,alickibook}, 
\begin{equation}
\frac{d\Xi _{t}}{dt}=-i[H,\Xi _{t}]_{-}+\sum_{i,j}a_{ij}(A_{i}\Xi
_{t}A_{j}^{\dagger }-\frac{1}{2}\{A_{j}^{\dagger }A_{i},\Xi _{t}\}_{+}).
\label{Bipa}
\end{equation}
Here, $[a,b]_{-}\equiv ab-ba$ and $\{a,b\}_{+}\equiv ab+ba$ denote a
commutator and anticommutator operations respectively. $H$ is an arbitrary
Hamiltonian operator, while $\{A_{i}\}$ is an arbitrary basis of operators
in $\mathcal{H}_{sc}.$ Finally, $\{a_{ij}\}$ is a Hermitian (positive
semi-definite) matrix of rate coefficients. The indexes $i$ and $j$ run from
one up to $(\dim \mathcal{H}_{sc})^{2},$ where $\dim \mathcal{H}_{sc}$ is
the dimension of $\mathcal{H}_{sc}.$

Eq.~(\ref{Bipa}) is the most general local-in-time evolution for the density
matrix $\Xi _{t}.$ Both subsystems are in general quantum ones. A hybrid
quantum-classical solution is obtained when the bipartite state, at
arbitrary times, can be written as 
\begin{equation}
\Xi _{t}=\sum_{c}\rho _{t}^{c}\otimes |c\rangle \langle c|.  \label{Suma}
\end{equation}
Here, $\{|c\rangle \}$ is an (fixed) orthogonal basis in $\mathcal{H}_{c},$ $
\langle c|c^{\prime }\rangle =\delta _{cc^{\prime }},$ which in turn
fulfills $\sum_{c}|c\rangle \langle c|=\mathrm{I}_{c}.$ With $\mathrm{I}$ we
denote the identity operator. $\rho _{t}^{c}$ are (conditional) quantum
states in $\mathcal{H}_{s}.$ Introducing the partial subsystems states, $
\rho _{t}\equiv \mathrm{Tr}_{c}[\Xi _{t}]$ and $\sigma _{t}\equiv \mathrm{Tr}
_{s}[\Xi _{t}],$ where $\mathrm{Tr}[\bullet ]$\ is the trace operation, it
follows the partial quantum state 
\begin{equation}
\rho _{t}=\sum_{c}\rho _{t}^{c},  \label{PartialStates}
\end{equation}
while for the classical subsystem we get 
\begin{equation}
\sigma _{t}=\sum_{c}p_{t}^{c}|c\rangle \langle c|,\ \ \ \ \ \ \ \ \ \ \
p_{t}^{c}\equiv \mathrm{Tr}_{s}[\rho _{t}^{c}].  \label{RhoClass}
\end{equation}
The state of the quantum subsystem $\rho _{t}$ is defined by the addition of
the unnormalized states $\{\rho _{t}^{c}\}.$ The state $\sigma _{t}$
corresponding to the classical subsystem is an incoherent statistical
mixture, where the weights (probabilities $p_{t}^{c}$) of each projector $
|c\rangle \langle c|$ is given by $\mathrm{Tr}_{s}[\rho _{t}^{c}].$ Thus,
classicality here means a system that is always incoherent in a given fixed
basis.

In general, the solution of Eq.~(\ref{Bipa}) cannot be written as a hybrid
state [Eq.~(\ref{Suma})]. Nevertheless, such kind of state becomes a
solution if one restrict the Hamiltonian $H,$ the operators $\{A_{i}\},$ and
the matrix of coefficients $\{a_{ij}\}.$ In fact, an appropriate choice of
these elements allows us to embed an arbitrary (completely positive) hybrid
time-evolution in the bipartite Hilbert space\ $\mathcal{H}_{sc}.$ The
bipartite initial condition must also be restricted. An uncorrelated state
is the most simple assumption, $\Xi _{0}=\rho _{0}\otimes \sigma _{0}=\rho
_{0}\otimes \sum_{c}p_{0}|c\rangle \langle c|,$ where the weights satisfy $
0\leq p_{0}\leq 1$ and $\sum_{c}p_{0}=1.$

Depending on the choice of the above elements different quantum-classical
coupling\ mechanisms emerge. For their specific formulation it is necessary
to introduce a basis of operators for each subsystem. For the quantum system
the basis is taken as $\mathrm{I}_{s}\cup \{V_{\mu }\},$ where $\mu
=1,\cdots ,(\dim \mathcal{H}_{s})^{2}-1.$ Notice that the identity operator $
\mathrm{I}_{s}$ is one element of this basis. Consequently, the set of
operators $\{V_{\mu }\}$ is traceless. For the \textquotedblleft
classical\textquotedblright\ subsystem the basis of operators is $
\{|c\rangle \langle c^{\prime }|\}$ where $c,c^{\prime }=1,\cdots \dim 
\mathcal{H}_{c}.$ In this case, the identity operator is obtained as $
\sum_{c}|c\rangle \langle c|=\mathrm{I}_{c}.$ By construction, the following
bipartite embedding guarantees that each coupling mechanism is independent
of the other ones.

\subsection{First case}

The first case we dealt with emerges after taking the operators 
\begin{equation}
A_{i}\rightarrow V_{\mu }\otimes |c\rangle \langle c|,\ \ \ \ \
H=\sum_{c}H_{c}\otimes |c\rangle \langle c|,  \label{FirstOperators}
\end{equation}
and the coefficients $a_{ij}\rightarrow a_{(\mu c)(\nu c^{\prime })}\delta
_{cc^{\prime }}\rightarrow \eta _{c}^{\mu \nu }.$ Inserting these elements
in Eq.~(\ref{Bipa}) the evolution of the quantum states $\{\rho _{t}^{c}\}$
can be written as 
\begin{equation}
\frac{d\rho _{t}^{c}}{dt}=-i[H_{c},\rho _{t}^{c}]_{-}+\sum_{\mu \nu }\eta
_{c}^{\mu \nu }(V_{\mu }\rho _{t}^{c}V_{\nu }^{\dagger }-\frac{1}{2}\{V_{\nu
}^{\dagger }V_{\mu },\rho _{t}^{c}\}_{+}).  \label{Uno}
\end{equation}
Here, $H_{c}$ is an arbitrary Hamiltonian parametrized by $c.$ The indexes $
\mu $ and $\nu $ run in the interval $1,\cdots (\dim \mathcal{H}_{s})^{2}-1$
(the identity matrix $\mathrm{I}_{s}$ is not included in the addition). The
coefficients $\{\eta _{c}^{\mu \nu }\}$ for each index $c$ must to define a
positive semi-definite matrix (with matrix indexes $\mu $ and $\nu $). This
property is inherited from the positivity constraint valid for $a_{ij}$ in
Eq.~(\ref{Bipa}).

The evolution~(\ref{Uno}) has a simple interpretation. The states $\{\rho
_{t}^{c}\}$ evolve independently of each other. Their dynamics is set by the
Hamiltonian $H_{c}$ and the matrix of rate coefficients $\eta _{c}^{\mu \nu
}.$ Both objects are parametrized by the state $(c)$ of the incoherent
subsystem. Thus, \textit{the evolution of the quantum subsystem depends on
the state of the classical one while the probabilities }$\{p_{t}^{c}\}$ 
\textit{\ are invariant ones. }There is not any backaction on the classical
subsystem. In fact, from Eq.~(\ref{Uno}) straightforwardly it follows $(d/dt)
\mathrm{Tr}_{s}[\rho _{t}^{c}]=0,$ which implies that 
\begin{equation}
\frac{d}{dt}\sigma _{t}=0.  \label{NoBackActionClass_1}
\end{equation}
Consequently, from Eq.~(\ref{PartialStates}) the system state $\rho _{t}$
can be read as a statistical mixture of the solutions of the Lindblad
dynamics~(\ref{Uno}). We notice that its time evolution, $(d/dt)\rho _{t},$
cannot be written in general as a time-independent Lindblad equation.

A self-independent dynamics of the quantum subsystem is covered by taking $
H_{c}\rightarrow H_{s}+H_{c}$ and $\eta _{c}^{\mu \nu }\rightarrow \eta
^{\mu \nu }+\eta _{c}^{\mu \nu }.$ Furthermore, by changing the basis of
operators $\{V_{\mu }\}$ it is always possible to obtain a diagonal matrix
of rate coefficients, $\eta _{c}^{\mu \nu }\rightarrow \delta _{\mu \nu
}\eta _{c}^{\mu }.$

\subsection{Second case}

The second case emerges by taking $H=0,$ 
\begin{equation}
A_{i}\rightarrow \mathrm{I}_{s}\otimes |c\rangle \langle c^{\prime }|,\ \ \
\ \ \ \ \ \ \ \ \ c\neq c^{\prime },  \label{SecondOperators}
\end{equation}
and $a_{ij}\rightarrow \delta _{ij}a_{i}\rightarrow \phi _{cc^{\prime }}.$
From Eq.~(\ref{Bipa}) it follows that 
\begin{equation}
\frac{d\rho _{t}^{c}}{dt}=\sum_{\substack{ c^{\prime }  \\ c^{\prime }\neq c 
}}\phi _{cc^{\prime }}\rho _{t}^{c^{\prime }}-\sum_{\substack{ c^{\prime } 
\\ c^{\prime }\neq c}}\phi _{c^{\prime }c}\rho _{t}^{c},  \label{Dos}
\end{equation}
where the rates fulfill $\phi _{cc^{\prime }}\geq 0$ $\forall c,c^{\prime }.$
In this case, the states $\{\rho _{t}^{c}\}$ are coupled between all of
them. In fact, Eq.~(\ref{Dos}) has the structure of a classical master
equation~\cite{kampen,gardiner,Risken} where the coupling terms do not
involve any information about the quantum system. Consequently, the
probabilities $p_{t}^{c}=\mathrm{Tr}_{s}[\rho _{t}^{c}]$ obey a classical
master equation 
\begin{equation}
\frac{dp_{t}^{c}}{dt}=\sum_{\substack{ c^{\prime }  \\ c^{\prime }\neq c}}
\phi _{cc^{\prime }}p_{t}^{c^{\prime }}-\sum_{\substack{ c^{\prime }  \\ 
c^{\prime }\neq c}}\phi _{c^{\prime }c}p_{t}^{c},  \label{ClassicalMaster2}
\end{equation}
defined by the rates $\{\phi _{c^{\prime }c}\}.$ In contrast, from Eq.~(\ref
{Dos}) it is simple to check that the state of the quantum subsystem 
[Eq.~(\ref{PartialStates})] does not evolve at all, 
\begin{equation}
\frac{d}{dt}\rho _{t}=0.  \label{QuantumFrozen_2}
\end{equation}
Thus, \textit{this mechanism endow the classical system with a proper
time-irreversible self dynamics,} and there is not any backaction on the
quantum subsystem. We notice that the combination of the first and second
cases allow to describe hybrid dynamics where the classical subsystem has an
independent self-evolution [Eqs.~(\ref{ClassicalMaster2})]\ whose state in
turn determine the (conditional) evolution of the quantum subsystem 
Eq.~(\ref{Uno})].

\subsection{Third case}

The third case emerges by taking $H=0,$ the operators 
\begin{equation}
A_{i}\rightarrow V_{\mu }\otimes |c\rangle \langle c^{\prime }|,\ \ \ \ \ \
c\neq c^{\prime },\ \ \ \ \ \ V_{\mu }\neq \mathrm{I}_{s},
\label{ThirdOperators}
\end{equation}
and $a_{ij}\rightarrow a_{(\mu cc^{\prime })(\nu \tilde{c}\tilde{c}^{\prime
})}\delta _{c\tilde{c}}\delta _{c^{\prime }\tilde{c}^{\prime }}\rightarrow
\gamma _{cc^{\prime }}^{\mu \nu }.$ From Eq.~(\ref{Bipa}) we get 
\begin{equation}
\frac{d\rho _{t}^{c}}{dt}=\sum_{\substack{ \mu ,\nu ,c^{\prime }  \\ 
c^{\prime }\neq c}}\gamma _{cc^{\prime }}^{\mu \nu }V_{\mu }\rho
_{t}^{c^{\prime }}V_{\nu }^{\dagger }-\frac{1}{2}\sum_{\substack{ \mu ,\nu
,c^{\prime }  \\ c^{\prime }\neq c}}\gamma _{c^{\prime }c}^{\mu \nu
}\{V_{\nu }^{\dagger }V_{\mu },\rho _{t}^{c}\}_{+},  \label{Tres}
\end{equation}
where the Hermitian matrices $\{\gamma _{c^{\prime }c}^{\mu \nu }\}$ (with
indexes $\mu $ and $\nu $) must be positive semi-definite $\forall
c,c^{\prime }.$ The evolution of the states $\{\rho _{t}^{c}\}$ has the
structure of a classical master equation. Nevertheless, the couplings are
mediated by operators of the quantum subsystem. Motivated by this property,
this kind of evolutions were named as Lindblad rate equations~\cite{rate}.

The underlying stochastic dynamics associated to Eq.~(\ref{Tres}) can be
read straightforwardly by considering the diagonal case $\gamma _{c^{\prime
}c}^{\mu \nu }=\delta ^{\mu \nu }\gamma _{c^{\prime }c}^{\mu }$ (this
property can always be achieved by changing the basis of operators $\{V_{\mu
}\}).$ Thus, we realize that whenever the quantum subsystem suffers a
transition induced by the operator $V_{\mu },$ that is, $\rho
_{t}^{c}\rightarrow V_{\mu }\rho _{t}^{c}V_{\mu }^{\dagger },$ the classical
subsystem simultaneously realize the (incoherent) transition $|c\rangle
\rightarrow |c^{\prime }\rangle .$ Therefore,\textit{\ both subsystems are
inherently coupled to each other.} In fact, a transition in one subsystem is
always endowed by a corresponding transition in the other subsystem. There
is a mutual backaction between both subsystems.

\subsubsection*{Coupling symmetries and backaction properties}

In general, due to their intrinsic coupling, the evolution of each part
cannot be written without involving information about the other. However,
there are two exceptions to this rule. The \textit{first} case emerges when
it is fulfilled the coupling symmetry property 
\begin{equation}
\gamma _{c^{\prime }c}^{\mu \nu }=\gamma _{c^{\prime }}^{\mu \nu }\ \ \ \ \
\ \forall \mu ,\nu ,c,c^{\prime }.  \label{Condition1}
\end{equation}
Hence, the transition rates do not depend on the \textquotedblleft
starting\textquotedblright\ state $|c\rangle .$ From Eqs.~(\ref
{PartialStates}) and~(\ref{Tres}) the state of the quantum subsystem evolves
as 
\begin{equation}
\frac{d\rho _{t}}{dt}=\sum_{\mu ,\nu }\gamma _{0}^{\mu \nu }(V_{\mu }\rho
_{t}V_{\nu }^{\dagger }-\frac{1}{2}\{V_{\nu }^{\dagger }V_{\mu },\rho
_{t}\}_{+}),  \label{QuantumInd}
\end{equation}
where $\gamma _{0}^{\mu \nu }\equiv \sum_{c^{\prime }}\gamma _{c^{\prime
}}^{\mu \nu }.$ This is a standard Lindblad equation that does not depends
on the classical subsystem, which implies that there is not any backaction
on the quantum subsystem. On the other hand, under the assumption~(\ref
{Condition1}) it is not possible to obtain a closed expression for $
(d/dt)p_{t}^{c}$ without involving information about the quantum subsystem.

The \textit{second }case occurs when it is fulfilled that 
\begin{equation}
\sum_{\mu ,\nu }\gamma _{c^{\prime }c}^{\mu \nu }V_{\nu }^{\dagger }V_{\mu
}=\gamma _{c^{\prime }c}^{0}\mathrm{I}_{s}\ \ \ \ \ \ \forall c,c^{\prime },
\label{Condition2}
\end{equation}
where $\{\gamma _{c^{\prime }c}^{0}\}$\ are constant (rates) of
proportionality. From Eqs.~(\ref{RhoClass}) and~(\ref{Tres}) the evolution
of the probabilities of the classical subsystem read 
\begin{equation}
\frac{dp_{t}^{c}}{dt}=\sum_{\substack{ c^{\prime }  \\ c^{\prime }\neq c}}
\gamma _{cc^{\prime }}^{0}p_{t}^{c^{\prime }}-\sum_{\substack{ c^{\prime } 
\\ c^{\prime }\neq c}}\gamma _{c^{\prime }c}^{0}p_{t}^{c}.  \label{popuInde}
\end{equation}
This classical master equation does not depend on the degrees of freedom of
the quantum system, which implies that there is none backaction on the
classical subsystem. On the other hand, it is not possible to write a closed
evolution for the quantum state $\rho _{t}.$

In general the condition~(\ref{Condition1}) neither condition~(\ref
{Condition2}) are fulfilled. Nevertheless, there are also dynamics where
both conditions are satisfied. Thus, in this case the hybrid dynamics
correlates both subsystems but their partial dynamics are independent of the
other [see Eqs.~(\ref{QuantumInd}) and~(\ref{popuInde})].

\subsection{Fourth case}

The fourth case can be read as a \textquotedblleft \textit{coherent
superposition\textquotedblright\ of the second and third cases} [Eqs.~(\ref
{SecondOperators}) and~(\ref{ThirdOperators}) respectively]. In Eq.~(\ref
{Bipa}) we take $H=0$ and 
\begin{equation}
A_{i}\rightarrow (a_{\mu }\mathrm{I}_{s}+b_{\mu }V_{\mu })\otimes |c\rangle
\langle c^{\prime }|,\ \ \ \ \ \ \ c\neq c^{\prime },\ \ V_{\mu }\neq 
\mathrm{I}_{s},  \label{FourthOperators}
\end{equation}
where $\{a_{\mu }\}$ and $\{b_{\mu }\}$ are arbitrary complex coefficients.
Thus, the conditions $a_{\mu }\neq 0,$ $b_{\mu }=0$ and alternatively $
a_{\mu }=0,$ $b_{\mu }\neq 0,$ recover the second and third cases
respectively. Under the replacements $a_{ij}\rightarrow a_{(\mu cc^{\prime
})(\nu \tilde{c}\tilde{c}^{\prime })}\delta _{c\tilde{c}}\delta _{c^{\prime
} \tilde{c}^{\prime }}\rightarrow \gamma _{cc^{\prime }}^{\mu \nu },$ from
Eq.~(\ref{Bipa}) we get 
\begin{equation}
\frac{d\rho _{t}^{c}}{dt}=\sum_{\substack{ (\mu ,\nu )\cup \mathrm{I}
,c^{\prime }  \\ c^{\prime }\neq c}}\lambda _{cc^{\prime }}^{\mu \nu }V_{\mu
}\rho _{t}^{c^{\prime }}V_{\nu }^{\dagger }-\frac{1}{2}\sum_{\substack{ (\mu
,\nu )\cup \mathrm{I},c^{\prime }  \\ c^{\prime }\neq c}}\lambda _{c^{\prime
}c}^{\mu \nu }\{V_{\nu }^{\dagger }V_{\mu },\rho _{t}^{c}\}_{+}.
\label{Cuatro}
\end{equation}
This equation is similar to Eq.~(\ref{Tres}). Nevertheless, here the
identity operator $\mathrm{I}_{s}$ is also included in the addition. In
fact, the coefficients $\mu $ and $\nu $ run in the intervals $1,\cdots
\lbrack (\dim \mathcal{H}_{s})^{2}-1]\cup \mathrm{I.}$ In addition, the
rates are 
\begin{subequations}
\label{StrucuturalFeedback}
\begin{eqnarray}
\lambda _{cc^{\prime }}^{\mu \nu } &=&b_{\mu }\gamma _{cc^{\prime }}^{\mu
\nu }b_{\nu }^{\ast },\ \ \ \ \ \ \ \ \ \ \lambda _{cc^{\prime }}^{\mathrm{
II }}=\sum_{\mu \nu }a_{\mu }\gamma _{cc^{\prime }}^{\mu \nu }a_{\nu }^{\ast
}, \\
\lambda _{cc^{\prime }}^{\mu \mathrm{I}} &=&\sum_{\nu }b_{\mu }\gamma
_{cc^{\prime }}^{\mu \nu }a_{\nu }^{\ast },\ \ \ \ \ \lambda _{cc^{\prime
}}^{\mathrm{I}\nu }=\sum_{\mu }a_{\mu }\gamma _{cc^{\prime }}^{\mu \nu
}b_{\nu }^{\ast },
\end{eqnarray}
where the Hermitian matrices $\{\gamma _{c^{\prime }c}^{\mu \nu }\}$
(indexes $\mu $ and $\nu )$ must be positive semi-definite $\forall
c,c^{\prime }.$

By using that $\{V,\rho \}_{+}=2V\rho -[V,\rho ]_{-}$ and $\{V^{\dagger
},\rho \}_{+}=2\rho V^{\dagger }+[V^{\dagger },\rho ]_{-},$ Eq.~(\ref{Cuatro}
) can be rewritten as 
\end{subequations}
\begin{eqnarray}
\!\!\!\!\frac{d\rho _{t}^{c}}{dt}\!\!\ &=&\!\!\!\sum_{\substack{ \mu ,\nu
,c^{\prime }  \\ c^{\prime }\neq c}}\lambda _{cc^{\prime }}^{\mu \nu }V_{\mu
}\rho _{t}^{c^{\prime }}V_{\nu }^{\dagger }-\frac{1}{2}\sum_{\substack{ \mu
,\nu ,c^{\prime }  \\ c^{\prime }\neq c}}\lambda _{c^{\prime }c}^{\mu \nu
}\{V_{\nu }^{\dagger }V_{\mu },\rho _{t}^{c}\}_{+}\ \ \ \ \   \notag \\
&&\!\!\!+\sum_{c^{\prime }\neq c}\lambda _{cc^{\prime }}^{\mathrm{II}}\rho
_{t}^{c^{\prime }}-\sum_{c^{\prime }\neq c}\lambda _{c^{\prime }c}^{\mathrm{
\ II}}\rho _{t}^{c}  \notag \\
&&-i[H_{c},\rho _{t}^{c}]_{-}  \label{CuatroBis} \\
&&\!\!\!+\sum_{\substack{ \mu ,c^{\prime }  \\ c^{\prime }\neq c}}\lambda
_{cc^{\prime }}^{\mu \mathrm{I}}V_{\mu }\rho _{t}^{c^{\prime }}-\sum 
_{\substack{ \mu ,c^{\prime }  \\ c^{\prime }\neq c}}\lambda _{c^{\prime
}c}^{\mu \mathrm{I}}V_{\mu }\rho _{t}^{c}  \notag \\
&&\!\!\!+\sum_{\substack{ \nu ,c^{\prime }  \\ c^{\prime }\neq c}}\lambda
_{cc^{\prime }}^{\mathrm{I}\nu }\rho _{t}^{c^{\prime }}V_{\nu }^{\dagger
}-\sum_{\substack{ \nu ,c^{\prime }  \\ c^{\prime }\neq c}}\lambda
_{c^{\prime }c}^{\mathrm{I}\nu }\rho _{t}^{c}V_{\nu }^{\dagger },  \notag
\end{eqnarray}
where the Hamiltonian $H_{c}$\ reads 
\begin{equation}
H_{c}=\frac{i}{2}\sum_{\substack{ \mu ,c^{\prime }  \\ c^{\prime }\neq c}}
(\lambda _{c^{\prime }c}^{\mu \mathrm{I}}V_{\mu }-\lambda _{c^{\prime }c}^{ 
\mathrm{I}\mu }V_{\mu }^{\dagger }).
\end{equation}

In Eq.~(\ref{CuatroBis}), the contributions given by the first two lines are
equivalent to the couplings defined by Eqs.~(\ref{Dos}) and~(\ref{Tres}).
The remaining three lines are the extra coupling terms that define the
fourth case. The physical mechanism introduced by the vector $\{\lambda
_{cc^{\prime }}^{\mathrm{I}\mu }\}$ can be related to a feedback mechanism~ 
\cite{wisemanBook} involving both the quantum and classical subsystems. On
the other hand, we notice that the Hamiltonians $\{H_{c}\}$ cannot be chosen
freely. They are part of the same coupling mechanism.

\subsubsection*{Coupling symmetries and backaction properties}

Similarly to the third case, the quantum and classical subsystems are
inherently coupled. Their time-evolution cannot be written without involving
information about the other part. Nevertheless, the situation defined by
Eq.~(\ref{Condition1}) can be extended to this case. If in Eq.~(\ref{Cuatro}
) the coupling rates fulfill the symmetry 
\begin{equation}
\lambda _{c^{\prime }c}^{\mu \nu }=\lambda _{c^{\prime }}^{\mu \nu }\ \ \ \
\ \ \forall (\mu ,\nu )\cup \mathrm{I},c,c^{\prime },
\end{equation}
the evolution of the quantum subsystem can be written as a Lindblad equation
with an extra Hamiltonian contribution 
\begin{equation}
\frac{d\rho _{t}}{dt}=-i[H_{0},\rho _{t}]_{-}+\sum_{\mu ,\nu }\lambda
_{0}^{\mu \nu }(V_{\mu }\rho _{t}V_{\nu }^{\dagger }-\frac{1}{2}\{V_{\nu
}^{\dagger }V_{\mu },\rho _{t}\}_{+}),  \label{QuantumIndependent}
\end{equation}
where $\lambda _{0}^{\mu \nu }=\sum_{c}\lambda _{cc^{\prime }}^{\mu \nu
}=\sum_{c}\lambda _{c}^{\mu \nu }$ and $H_{0}=(i/2)\sum_{\mu ,c^{\prime
}}(\lambda _{c^{\prime }}^{\mu \mathrm{I}}V_{\mu }-\lambda _{c^{\prime }}^{ 
\mathrm{I}\mu }V_{\mu }^{\dagger }).$ The above Lindblad structure follows
straightforwardly from Eq.~(\ref{CuatroBis}). On the other hand, the
situation defined by Eqs.~(\ref{Condition2}) and~(\ref{popuInde}) cannot be
extended to this case [Eq.~(\ref{Cuatro})].

\subsection{General case}

Any quantum-classical hybrid evolution can be written as a combination of
the above four coupling mechanisms, that is, an addition of the evolutions
defined by Eqs.~(\ref{Uno}), (\ref{Dos}), (\ref{Tres}), and~(\ref{Cuatro}).
The argument that support this result is that the above choice of bipartite
operators [Eqs.~(\ref{FirstOperators}), (\ref{SecondOperators}), ( \ref
{ThirdOperators}), and~(\ref{FourthOperators})] cover all possible cases. As
a matter of fact, here we analyze the case in which, for simplicity, $H=0$
[the general form of a unitary evolution is envisaged by Eq.~(\ref
{FirstOperators})] and 
\begin{equation}
A_{i}\rightarrow \mathrm{I}_{s}\otimes |c\rangle \langle c^{\prime }|,\ \ \
\ \ \ A_{i}\rightarrow V_{\mu }\otimes |c\rangle \langle c^{\prime }|,\ \ \
\ \ \ c\neq c^{\prime },\ \ V_{\mu }\neq \mathrm{I}_{s},
\end{equation}
while $a_{ij}\rightarrow a_{(\mu cc^{\prime })(\nu \tilde{c}\tilde{c}
^{\prime })}\delta _{c\tilde{c}}\delta _{c^{\prime }\tilde{c}^{\prime
}}\rightarrow \lambda _{cc^{\prime }}^{\mu \nu }.$ Notice that the operators 
$\mathrm{I}_{s}$ and $V_{\mu }$ are taken separately. After similar
calculations steps one arrive to the evolution defined by Eq.~(\ref{Cuatro}
). Nevertheless, here the matrix of rate coefficients defined by the
parameters $\lambda _{cc^{\prime }}^{\mu \nu }$ do not fulfill the
relations~(\ref{StrucuturalFeedback}). In fact, the unique constraint on
these parameters is a positivity condition inherited from the underlying
Lindblad structure, Eq.~(\ref{Bipa}).

In order to analyze the consequences of the positivity constraint we
explicitly write the matrix of rate coefficients as 
\begin{equation}
\left( 
\begin{array}{cc}
\{\lambda _{cc^{\prime }}^{\mu \nu }\} & \lambda _{cc^{\prime }}^{\mu 
\mathrm{I}} \\ 
\lambda _{cc^{\prime }}^{\mathrm{I}\nu } & \lambda _{cc^{\prime }}^{\mathrm{
\ II}}
\end{array}
\right) =\left( 
\begin{array}{cccc}
\lambda _{cc^{\prime }}^{11} & \lambda _{cc^{\prime }}^{12} & \cdots & 
\lambda _{cc^{\prime }}^{1\mathrm{I}} \\ 
\lambda _{cc^{\prime }}^{21} & \lambda _{cc^{\prime }}^{22} & \cdots & 
\lambda _{cc^{\prime }}^{2\mathrm{I}} \\ 
\vdots & \vdots & \ddots & \vdots \\ 
\lambda _{cc^{\prime }}^{\mathrm{I}1} & \lambda _{cc^{\prime }}^{\mathrm{I}2}
& \cdots & \lambda _{cc^{\prime }}^{\mathrm{II}}
\end{array}
\right) .  \label{Matrizon}
\end{equation}
This matrix (whose elements are labelled by indexes $\mu ,$ $\nu ,\ $and $
\mathrm{I}$) must be positive definite for all possible $c,c^{\prime }$ $
(c\neq c^{\prime }).$ The rate defined by $\lambda _{cc^{\prime }}^{\mathrm{
\ II}}$ has the same role than in Eq.~(\ref{Dos}). The submatrix defined by
the coefficients $\{\lambda _{cc^{\prime }}^{\mu \nu }\}$ has the same role
than in Eq.~(\ref{Tres}). The vectors (for each $c,c^{\prime })\ \lambda
_{cc^{\prime }}^{\mu \mathrm{I}}=(\lambda _{cc^{\prime }}^{\mathrm{I}\mu
})^{\ast }$ define extra coupling terms that are similar to those introduced
in Eq.~(\ref{Cuatro}).

Given that all minor matrices must also be positive definite (Sylvester's
criterion)~\cite{AlgebraMatrices} it follows that the submatrix $\{\lambda
_{cc^{\prime }}^{\mu \nu }\}$ must be positive definite and also that $
\lambda _{cc^{\prime }}^{\mathrm{II}}>0.$ Hence, the positivity of the
matrix~(\ref{Matrizon}) can be reduced to a condition on the vector$\
\{\lambda _{cc^{\prime }}^{\mu \mathrm{I}}\}$ (whose components are labeled
by index $\mu ).$ First, we analyze the case in which $\{\lambda
_{cc^{\prime }}^{\mu \nu }\}$ is a diagonal matrix. This property can always
be achieved by a proper choice of operators $\{V_{\mu }\}.$ In this
situation, it is possible to demonstrate that all submatrices of Eq.~(\ref
{Matrizon}) are positive definite (equivalent to positivity of a Schur
complement~\cite{AlgebraMatrices}) if 
\begin{equation}
\lambda _{cc^{\prime }}^{\mu \nu }=\delta _{u\nu }\lambda _{cc^{\prime
}}^{\mu }\ \ \ \ \Rightarrow \ \ \ \ \sum_{\mu \subset \mathrm{M}}\ \ \frac{
|\lambda _{cc^{\prime }}^{\mu \mathrm{I}}|^{2}}{(\lambda _{cc^{\prime
}}^{\mu }\lambda _{cc^{\prime }}^{\mathrm{II}})}\leq 1\ \ \ \ \ \forall
(c,c^{\prime }),  \label{diagonal}
\end{equation}
where $\mathrm{M}$ is an arbitrary subset of the possible $\mu $-values\
that define the corresponding minor matrix. The most stringent condition,
that in turn cover all conditions, is when $\mathrm{M}=\{1,\cdots (\dim 
\mathcal{H}_{s})^{2}-1\}.$ In this case, Eq.~(\ref{diagonal}) is equivalent
to the positivity of the determinant of the matrix~(\ref{Matrizon}) (with $
\lambda _{cc^{\prime }}^{\mu \nu }=\delta _{u\nu }\lambda _{cc^{\prime
}}^{\mu }).$

From Eq.~(\ref{diagonal}) we notice that the allowed \textquotedblleft
space\textquotedblright\ for the complex vector $\{\lambda _{cc^{\prime
}}^{\mu \mathrm{I}}\}$ is defined by the interior of a multidimensional
complex ellipse whose parameters are $(\lambda _{cc^{\prime }}^{\mu }\lambda
_{cc^{\prime }}^{\mathrm{II}})$ (notice that $\lambda _{cc^{\prime }}^{\mu
}>0$ and $\lambda _{cc^{\prime }}^{\mathrm{II}}>0).$ Using that $|\lambda
_{cc^{\prime }}^{\mu \mathrm{I}}|^{2}=\lambda _{cc^{\prime }}^{\mathrm{I}\mu
}\lambda _{cc^{\prime }}^{\mu \mathrm{I}},$ from Eq.~(\ref{diagonal}) the
positivity of the rate coefficients [matrix~(\ref{Matrizon})] is guaranteed 
\textit{in general} if 
\begin{equation}
\sum_{\mu \nu }\lambda _{cc^{\prime }}^{\mathrm{I}\nu }[\lambda _{cc^{\prime
}}^{-1}]^{\nu \mu }\lambda _{cc^{\prime }}^{\mu \mathrm{I}}\leq \lambda
_{cc^{\prime }}^{\mathrm{II}}\ \ \ \ \ \ \forall (c,c^{\prime }),
\label{ConstraintPositivity}
\end{equation}
where $[\lambda _{cc^{\prime }}^{-1}]^{\nu \mu }$ are the components of the
inverse matrix defined by $\{\lambda _{cc^{\prime }}^{\nu \mu }\}$ (which in
the general case is non-diagonal). Similarly to the diagonal case [Eq.~(\ref
{diagonal})], given $\{\lambda _{cc^{\prime }}^{\mu \nu }\}$ and $\lambda
_{cc^{\prime }}^{\mathrm{II}}$ the expression~(\ref{ConstraintPositivity})
defines a constraint to be fulfilled by the complex vector $\{\lambda
_{cc^{\prime }}^{\mathrm{I}\mu }\}.$

It is simple to check that the matrix of the fourth coupling mechanism
[Eq.~( \ref{StrucuturalFeedback})] satisfies the equality defined by Eq.~(
\ref{ConstraintPositivity}). Equivalently, in the diagonal version, the
fourth mechanism corresponds to the external boundary of the ellipsoid
defined by Eq.~(\ref{diagonal}). Thus, given an arbitrary evolution defined
by the positive definite matrix~(\ref{Matrizon}) it can always be written as
a combination of the fourth mechanism and extra contributions of the second
and third coupling mechanisms. These last contributions induce the
inequality defined by Eq.~(\ref{ConstraintPositivity}) [equivalently the
interior of the ellipsoid~(\ref{diagonal})]. We notice that the previous
decomposition is always possible because it implies to writing a positive
definite matrix [Eq.~(\ref{Matrizon})] in terms of other positive definite
matrices [matrices of rate coefficients associated to Eq.~(\ref{Dos}), (\ref
{Tres}), and~(\ref{Cuatro})].

\section{Diffusive Approximation}

The mechanisms characterized in the previous section are independent. Each
can be added freely to define a specific hybrid dynamics. Here, we study the
situation in which the incoherent classical subsystem can be labeled by a
continuous variable. Basically, this assumption implies the replacements $
c\rightarrow q$ and $\sum_{c}\rightarrow \int dq,$ where $q$ is a real and
continuous coordinate that label the state of the classical subsystem.
Furthermore, under the replacement $\rho _{c}\rightarrow \varrho _{t}(q)$
the hybrid state [Eq.~(\ref{Suma})] is written as 
\begin{equation}
\Xi _{t}=\int_{-\infty }^{+\infty }dq\varrho _{t}(q)\otimes |q\rangle
\langle q|.  \label{HybridStateContinuo}
\end{equation}
Consequently, the partial states [Eqs.~(\ref{PartialStates}) and~(\ref
{RhoClass})] become 
\begin{equation}
\rho _{t}=\int_{-\infty }^{+\infty }dq\varrho _{t}(q),\ \ \ \ \ \ \sigma
_{t}=\int_{-\infty }^{+\infty }dq\mathrm{Tr}_{s}[\varrho _{t}(q)]|q\rangle
\langle q|.
\end{equation}
Notice that $\varrho _{t}(q)$ can be read as a density (in the classical
space with coordinate $q$) of the state of the quantum subsystem. In fact, $
dq\varrho _{t}(q)$ is the density matrix operator when the classical
coordinate is \textquotedblleft projected\textquotedblright\ into the
interval $(q,q+dq).$

Under a straightforward change of notation all previous hybrid models can be
extended to the present situation. The question we dealt out now is when the
time-evolution of the \textquotedblleft state\textquotedblright\ $\varrho
_{t}(q)$ approach a \textit{diffusive limit}. Specifically, this means that $
(\partial /\partial t)\varrho _{t}(q)$ can be approached in terms of the
first and second \textquotedblleft spatial\textquotedblright\ derivatives $
(\partial /\partial q)\varrho _{t}(q)$ and $(\partial /\partial
q)^{2}\varrho _{t}(q)$ respectively.

\subsection{First case}

For the first coupling mechanism [Eq.~(\ref{Uno})], under the replacements $
\rho _{t}^{c}\rightarrow \varrho _{t}(q),$ $\eta _{c}^{\mu \nu }\rightarrow
\eta ^{\mu \nu }(q),$ the time evolution in the continuous case becomes 
\begin{eqnarray}
\frac{\partial \varrho _{t}(q)}{\partial t} &=&-i[H(q),\varrho
_{t}(q)]_{-}+\sum_{\mu \nu }\eta ^{\mu \nu }(q)V_{\mu }\varrho _{t}(q)V_{\nu
}^{\dagger }  \notag \\
&&-\sum_{\mu \nu }\eta ^{\mu \nu }(q)\frac{1}{2}\{V_{\nu }^{\dagger }V_{\mu
},\varrho _{t}(q)\}_{+}.  \label{UnoContinuo}
\end{eqnarray}
Here, the matrix of rate coefficients $\eta ^{\mu \nu }(q)$ must be positive
definite $\forall q.$ Given that the states$\{\varrho _{t}(q)\}$ are not
coupled to each other, the time evolution does not involve any coordinate
partial derivative.

\subsection{Second case}

For the continuos limit of the second case [Eq.~(\ref{Dos})], the associated
rates are rewritten as $\phi _{cc^{\prime }}\rightarrow \phi (q,q^{\prime
})\rightarrow \phi (q-q^{\prime }|q^{\prime }).$ This last object gives the
rate for a \textquotedblleft jump\textquotedblright\ of size $(q-q^{\prime
}) $ \textit{given} that the starting coordinate is $q^{\prime }.$ Under the
change of variables $r=q-q^{\prime }$ it follows 
\begin{equation}
\frac{\partial \varrho _{t}(q)}{\partial t}=\int dr\phi (r|q-r)\varrho
_{t}(q-r)-\int dr\phi (r|q)\varrho _{t}(q).  \label{PreDifusiva}
\end{equation}
This expression provides the basis for establishing the assumptions under
which a diffusive approximation applies~\cite{kampen}. (i) The \textit{first
assumption }is that only \textquotedblleft small jumps\textquotedblright\
occur, that is, $\phi (r|q)$ is a sharply peaked function of $r$\ but varies
slowly with $q.$ (ii) The \textit{second assumption} is that the solution $
\varrho _{t}(q)$ [which obeys Eq.~(\ref{PreDifusiva})] also varies slowly
with $q.$ This last assumption must be valid in \textit{any basis associated
to the Hilbert space of the quantum subsystem}. Therefore, in Eq.~(\ref
{PreDifusiva}) it is possible to deal with the shift from $q$ to $q-r$ by
means of a (second order) Taylor expansion 
\begin{equation}
\phi (r|q-r)\varrho _{t}(q-r)\simeq \Big{(}1-r\frac{\partial }{\partial q}+ 
\frac{r^{2}}{2}\frac{\partial ^{2}}{\partial q^{2}}\Big{)}\lbrack \phi
(r|q)\varrho _{t}(q)],  \label{Taylor}
\end{equation}
which delivers the \textit{diffusive approximation} 
\begin{equation}
\frac{\partial \varrho _{t}(q)}{\partial t}=-\frac{\partial }{\partial q}
[\Phi _{1}(q)\varrho _{t}(q)]+\frac{1}{2}\frac{\partial ^{2}}{\partial q^{2}}
[\Phi _{2}(q)\varrho _{t}(q)].  \label{DosDifusiva}
\end{equation}
In here, the \textquotedblleft jump moments\textquotedblright\ are 
\begin{equation}
\Phi _{m}(q)\equiv \int dr\phi (r|q)r^{m}.  \label{2JumpMoments}
\end{equation}

The \textquotedblleft Fokker-Planck structure\textquotedblright\ ~(\ref
{DosDifusiva}) is an approximation to the original master equation~(\ref
{PreDifusiva}). It is relevant to enforce that it is valid in both a
coarse-grained time-scale $\Delta t$ and coarse-grained coordinate-scale $
\Delta q,$ 
\begin{equation}
\Delta t\equiv 1/\Phi _{0},\ \ \ \ \ \ \Delta q\equiv \sqrt{\Phi _{2}/(2\Phi
_{0})}.
\end{equation}
The parameter $\Delta t$ defines the average time between consecutive jumps
of the classical coordinate while $\Delta q$ measures their size (for
simplicity the dependence with $q$ here is omitted). In particular, at the 
\textit{initial stage} Eq.~(\ref{DosDifusiva}) provides an approximation to
Eq.~(\ref{PreDifusiva}) if 
\begin{equation}
\Phi _{0}t\gg 1,\ \ \ \ \ \ \ \ \ \ \ \ \ |q|\Big{(}\frac{\Phi _{2}}{2\Phi
_{0}}\Big{)}^{-1/2}\gg 1.  \label{ShortScales}
\end{equation}
In fact, the diffusive approximation becomes valid only after several jump
processes have occurred. The second inequality relies on assuming a
coordinate initial condition concentrated in $q=0.$ On the other hand, we
remark that, independently of which initial conditions are considered, 
\textit{here the diffusive approximation guarantees that }$\varrho _{t}(q)$ 
\textit{\ is a physical state even in the short time and small length scales}
[complementary regime to that defined by Eq. (\ref{ShortScales})]. While
Eq.~(\ref{DosDifusiva}) does not depend explicitly on $\Phi _{0},$ this
time-scale becomes relevant in the next cases.

In order to clarify the underlying assumptions of the above diffusive
approximation we introduce an explicit example of transition rate, 
\begin{equation}
\phi (r|q)=\Big{[}\frac{1}{\tau _{0}}+\frac{1}{\delta \tau _{0}}\mathrm{sgn}
(r)\Big{]}\Big{(}\frac{1}{2r_{0}}e^{-|r|/r_{0}}\Big{)}.  \label{ExampleRate}
\end{equation}
Here, $\tau _{0}>0$ and $|\delta \tau _{0}|\leq \tau _{0}$ are
characteristic times, while the characteristic length fulfills $r_{0}>0.$
The sign function is defined as $\mathrm{sgn}(r)=\pm 1$ for $r\gtrless 0.$
The relevant moments [Eq.~(\ref{2JumpMoments})]\ that define the diffusive
approximation are~\cite{pie} 
\begin{equation}
\Phi _{0}=\frac{1}{\tau _{0}},\ \ \ \ \ \Phi _{1}=\frac{r_{0}}{\delta \tau
_{0}},\ \ \ \ \ \Phi _{2}=\frac{2r_{0}^{2}}{\tau _{0}}.
\end{equation}
With these expressions the approximation~(\ref{DosDifusiva}) can also be
obtained under the formal limits $\tau _{0}\rightarrow 0,\ r_{0}\rightarrow
0,\ 2r_{0}^{2}/\tau _{0}\rightarrow \Phi _{2},$ jointly with $r_{0}/\delta
\tau _{0}\rightarrow \Phi _{1}$~\cite{kampen}.

\subsection{Third case}

For simplifying the presentation, the third case is studied in its diagonal
form. Hence, in Eq.~(\ref{Tres}) we take $\gamma _{cc^{\prime }}^{\mu \nu
}=\delta _{\mu \nu }\gamma _{cc^{\prime }}^{\mu }.$ Similarly to the
previous case, under the replacements $\gamma _{cc^{\prime }}^{\mu
}\rightarrow \gamma ^{\mu }(q,q^{\prime })\rightarrow \gamma ^{\mu
}(q-q^{\prime }|q^{\prime }),$ and the change of variable $r=q-q^{\prime }$
we get 
\begin{eqnarray}
\frac{\partial \varrho _{t}(q)}{\partial t} &=&+\int dr\gamma ^{\mu
}(r|q-r)V_{\mu }\varrho _{t}(q-r)V_{\mu }^{\dagger }  \notag \\
&&-\frac{1}{2}\int dr\gamma ^{\mu }(r|q)\{V_{\mu }^{\dagger }V_{\mu
},\varrho _{t}(q)\}_{+}.  \label{TerceroContinuo}
\end{eqnarray}
For shorten the expression, a sum symbol $\sum_{\mu }$ is omitted on the
right side. The same notation is used in the following expressions and
sections.

While the structure of Eq.~(\ref{TerceroContinuo}) seems to be similar to
that of Eq.~(\ref{PreDifusiva}) they differ radically. In order to show this
feature, first we introduce the jump moments 
\begin{equation}
\Gamma _{m}^{\mu }(q)\equiv \int dr\gamma ^{\mu }(r|q)r^{m},
\end{equation}
which allow us to rewrite Eq.~(\ref{TerceroContinuo}) as 
\begin{eqnarray}
\frac{\partial \varrho _{t}(q)}{\partial t} &=&\Gamma _{0}^{\mu }(q)(V_{\mu
}\varrho _{t}(q)V_{\mu }^{\dagger }-\frac{1}{2}\{V_{\mu }^{\dagger }V_{\mu
},\varrho _{t}(q)\}_{+})  \notag \\
&&+V_{\mu }\left\{ \int dr\gamma ^{\mu }(r|q-r)\varrho _{t}(q-r)\right. 
\notag \\
&&-\left. \int dr\gamma ^{\mu }(r|q)\varrho _{t}(q)\right\} V_{\mu
}^{\dagger }.  \label{PTresSeparada}
\end{eqnarray}
In contrast to Eq.~(\ref{PreDifusiva}), here the first line on the right
side consists of a Lindblad contribution whose rate is $\Gamma _{0}^{\mu
}(q).$ Consistently with a diffusive approximation, this parameter, $\Gamma
_{0}^{\mu }(q),$ should define the minor time scale of the problem.

The second and third lines, $\int dr\gamma ^{\mu }(r|q-r)\varrho
_{t}(q-r)-\int dr\gamma ^{\mu }(r|q)\varrho _{t}(q),$ have the same
structure than Eq.~(\ref{PreDifusiva}). Nevertheless, these contributions
are under the action of the superoperator $V_{\mu }\{\bullet \}V_{\mu
}^{\dagger }.$ Given that the operators $\{V_{\mu }\}$ do not include the
identity matrix, this superoperator couples the discrete states of the
quantum system, inducing, in general, underlying finite transformations of
the state $\varrho _{t}(q).$ Consequently, even when the jump rates $
\{\gamma ^{\mu }(r|q)\}$ are sharply peaked functions of $r$\ (first
assumption), the solution $\varrho _{t}(q)$ [which obeys Eq.~(\ref
{PTresSeparada})] does not varies slowly with $q$ (second condition does not
applies). In general, the diffusive approximation [Eq.~(\ref{Taylor})] may
only applies in a given \textquotedblleft direction\textquotedblright\ in
the Hilbert space (defined by the eigenvectors of $V_{\mu }$) but it cannot
be applied to arbitrary directions. We conclude that \textit{a diffusive
approximation cannot be applied to the third case.}

\subsubsection*{Combined second and third cases}

A diffusive approximation can be applied in we consider the combined action
of the dynamics corresponding to the second and third coupling mechanisms.
From Eqs.~(\ref{PreDifusiva}) and~(\ref{PTresSeparada}) we get 
\begin{eqnarray}
\frac{\partial \varrho _{t}(q)}{\partial t} &\approx &\Gamma _{0}^{\mu
}(q)(V_{\mu }\varrho _{t}(q)V_{\mu }^{\dagger }-\frac{1}{2}\{V_{\mu
}^{\dagger }V_{\mu },\varrho _{t}(q)\}_{+})  \notag \\
&&\!\!\!\!\!\!\!\!\!-\frac{\partial }{\partial q}[\Phi _{1}(q)\varrho
_{t}(q)]+\frac{1}{2}\frac{\partial ^{2}}{\partial q^{2}}[\Phi _{2}(q)\varrho
_{t}(q)]  \label{ThirdSecond} \\
&&\!\!\!\!\!\!\!\!\!+V_{\mu }\Big{\{}-\frac{\partial }{\partial q}[\Gamma
_{1}^{\mu }(q)\varrho _{t}(q)]+\frac{1}{2}\frac{\partial ^{2}}{\partial q^{2}
}[\Gamma _{2}^{\mu }(q)\varrho _{t}(q)]\Big{\}}V_{\mu }^{\dagger }.  \notag
\end{eqnarray}
\textit{This approximation applies when the second coupling mechanism is
dominant}. Thus, it must be fulfilled that 
\begin{equation}
\Gamma _{0}^{\mu }(q)\neq 0,\ \ \ \ \ \ \ \Gamma _{m}^{\mu }(q)\ll \Phi
_{m}(q)\ \ \ \ \ \ \forall \mu ,q,m.  \label{ThirdConstraints}
\end{equation}
The condition $\Gamma _{0}^{\mu }(q)\neq 0$ inherently arises from the third
case. On the other hand, the inequalities between the jump moments, $\Gamma
_{m}^{\mu }(q)\ll \Phi _{m}(q),$ guaranty that the second coupling mechanism
is the dominant one, allowing in consequence a diffusive approximation (in
any possible direction of the quantum subsystem Hilbert space). In addition,
the approximation~(\ref{ThirdSecond}) is valid under the scale constraints~(
\ref{ShortScales}). In contrast to the second case [Eq.~(\ref{DosDifusiva}
)], here it is not possible to guaranty that $\varrho _{t}(q)$ is a physical
state (positive definite) in the short time and small length scales. This
condition is recovered in the limit $\Gamma _{m}^{\mu }(q)/\Phi
_{m}(q)\rightarrow 0,$ which from Eq.~(\ref{ThirdSecond}) delivers 
\begin{eqnarray}
\frac{\partial \varrho _{t}(q)}{\partial t} &=&\Gamma _{0}^{\mu }(q)(V_{\mu
}\varrho _{t}(q)V_{\mu }^{\dagger }-\frac{1}{2}\{V_{\mu }^{\dagger }V_{\mu
},\varrho _{t}(q)\}_{+})  \notag \\
&&\!\!\!-\frac{\partial }{\partial q}[\Phi _{1}(q)\varrho _{t}(q)]+\frac{1}{2
}\frac{\partial ^{2}}{\partial q^{2}}[\Phi _{2}(q)\varrho _{t}(q)].\ \ \ \ 
\label{DosTresDifusiva}
\end{eqnarray}
In this expression, any contribution related to third case is taken into
account solely by the first line, which has the structure of a Lindblad
equation. While Eq.~(\ref{DosTresDifusiva}) preserves the positivity of $
\varrho _{t}(q)$ notice that there is not any backaction (related to the
third case) on the classical subsystem. In fact, the evolution of the
probability density $\mathrm{Tr}_{s}[\varrho _{t}(q)]$ is governed by a
classical Fokker-Planck equation associated to the second mechanism.

\subsection{Fourth case}

Similarly to third case, in order to simplify the expressions in this case
[Eq.~(\ref{Cuatro})]\ we assume a diagonal matrix of rate coefficients, $
\lambda _{cc^{\prime }}^{\mu \nu }=\delta _{\mu \nu }\lambda _{cc^{\prime
}}^{\mu },$ which in terms of the relations~(\ref{StrucuturalFeedback})
implies\ that $\gamma _{cc^{\prime }}^{\mu \nu }\rightarrow \delta ^{\mu \nu
}\gamma _{cc^{\prime }}^{\mu }\rightarrow \delta ^{\mu \nu }\gamma ^{\mu
}(r|q).$ The jump rates becomes $\lambda _{cc^{\prime }}^{\mathrm{II}
}\rightarrow \lambda ^{\mathrm{II}}(r|q),$ $\lambda _{cc^{\prime }}^{\mu
}\rightarrow \lambda ^{\mu }(r|q),$ leading to the expressions 
\begin{equation}
\lambda ^{\mu }(r|q)=|b_{\mu }|^{2}\gamma ^{\mu }(r|q),\ \ \ \ \ \lambda ^{ 
\mathrm{II}}(r|q)=\sum_{\mu }|a_{\mu }|^{2}\gamma ^{\mu }(r|q).
\label{Lamdba_1}
\end{equation}
Similarly, under the notational replacement $\lambda _{c^{\prime }c}^{\mu 
\mathrm{I}}\rightarrow \lambda ^{\mu \mathrm{I}}(r|q)$ and $\lambda
_{c^{\prime }c}^{\mathrm{I}\nu }\rightarrow \lambda ^{\mathrm{I}\mu }(r|q),$
it follows 
\begin{equation}
\lambda ^{\mu \mathrm{I}}(r|q)=b_{\mu }\gamma ^{\mu }(r|q)a_{\mu }^{\ast },\
\ \ \ \ \lambda ^{\mathrm{I}\mu }(r|q)=a_{\mu }\gamma ^{\mu }(r|q)b_{\mu
}^{\ast }.  \label{Lamdba_2}
\end{equation}
In the above expressions $\gamma ^{\mu }(r|q)$ is a positive jumping rate
while $a_{\mu }$ and $b_{\mu }$ define the bipartite coupling operator~(\ref
{FourthOperators}).

Starting from the equivalent expression~(\ref{CuatroBis}), and assuming that
the rate $\lambda ^{\mathrm{II}}(r|q)$ is dominant over the processes
induced by $\lambda ^{\mu }(r|q),$ $\lambda ^{\mu \mathrm{I}}(r|q),$ and $
\rightarrow \lambda ^{\mathrm{I}\mu }(r|q)=\lambda ^{\mu }(r|q)^{\ast },$ it
is simple to arrive to the diffusive approximation [see Appendix] 
\begin{eqnarray}
\frac{\partial \varrho _{t}(q)}{\partial t} &\approx &\Lambda _{0}^{\mu
}(q)(V_{\mu }\varrho _{t}(q)V_{\mu }^{\dagger }-\frac{1}{2}\{V_{\mu
}^{\dagger }V_{\mu },\varrho _{t}(q)\}_{+})  \notag \\
&&\!\!\!\!\!\!\!\!\!+V_{\mu }\Big{\{}-\frac{\partial }{\partial q}[\Lambda
_{1}^{\mu }(q)\varrho _{t}(q)]+\frac{1}{2}\frac{\partial ^{2}}{\partial q^{2}
}[\Lambda _{2}^{\mu }(q)\varrho _{t}(q)]\Big{\}}V_{\mu }^{\dagger }  \notag
\\
&&\!\!\!\!\!\!\!\!\!-\frac{\partial }{\partial q}[\Lambda _{1}^{\mathrm{II}
}(q)\varrho _{t}(q)]+\frac{1}{2}\frac{\partial ^{2}}{\partial q^{2}}[\Lambda
_{2}^{\mathrm{II}}(q)\varrho _{t}(q)]  \notag \\
&&\!\!\!\!\!\!\!\!\!-i[H_{0}(q),\varrho _{t}(q)]_{-}  \label{FourthApproxy}
\\
&&\!\!\!\!\!\!\!\!\!\!\!+V_{\mu }\Big{\{}-\frac{\partial }{\partial q}
[\Lambda _{1}^{\mu \mathrm{I}}(q)\varrho _{t}(q)]+\frac{1}{2}\frac{\partial
^{2}}{\partial q^{2}}[\Lambda _{2}^{\mu \mathrm{I}}(q)\varrho _{t}(q)]
\Big{\}}  \notag \\
&&\!\!\!\!\!\!\!\!\!\!\!+\Big{\{}-\frac{\partial }{\partial q}[\Lambda _{1}^{
\mathrm{I}\mu }(q)\varrho _{t}(q)]+\frac{1}{2}\frac{\partial ^{2}}{\partial
q^{2}}[\Lambda _{2}^{\mathrm{I}\mu }(q)\varrho _{t}(q)]\Big{\}}V_{\mu
}^{\dagger },  \notag
\end{eqnarray}
where, for simplicity, on the right side terms that depend on index $\mu $\
we omitted a sum symbol $\sum_{\mu }.$ The Hamiltonian contribution is 
\begin{equation}
H_{0}(q)=\frac{i}{2}\sum_{\mu }(\Lambda _{0}^{\mu \mathrm{I}}(q)V_{\mu
}-\Lambda _{0}^{\mathrm{I}\mu }(q)V_{\mu }^{\dagger }),
\label{Hamiltonian_q}
\end{equation}
while the jump moments read 
\begin{equation}
\Lambda _{m}^{\mu }(q)\!\equiv \!\int \!dr\lambda ^{\mu }(r|q)r^{m},\ \ \
\Lambda _{m}^{\mathrm{II}}(q)\!\equiv \!\int \!dr\lambda ^{\mathrm{II}
}(r|q)r^{m},  \label{LambdaMom_1}
\end{equation}
and similarly for the nondiagonal contributions 
\begin{equation}
\Lambda _{m}^{\mu \mathrm{I}}(q)\!\equiv \!\int \!dr\lambda ^{\mu \mathrm{I}
}(r|q)r^{m},\!\ \ \!\Lambda _{m}^{\mathrm{I}\mu }(q)\!\equiv \!\int
\!dr\lambda ^{\mathrm{I}\mu }(r|q)r^{m}.\   \label{LambdaMom_2}
\end{equation}

The approximation~(\ref{FourthApproxy}) relies on assuming that the
diffusive process induced by $\lambda ^{\mathrm{II}}(r|q),$ which does not
depend on the Hilbert space orientation, is dominant over the processes
induced by $\lambda ^{\mu }(r|q)$ and $\lambda ^{\mu \mathrm{I}}(r|q)$
(notice that $\lambda ^{\mathrm{I}\mu }=[\lambda ^{\mu \mathrm{I}
}(r|q)]^{\ast }).$ Consequently, Eq.~(\ref{FourthApproxy}) approximates the
original evolution [Eq.~(\ref{CuatroBis}) or equivalently Eq.~(\ref
{CuartoMasterSplit})] if the jump moments satisfy 
\begin{equation}
\Lambda _{m}^{\mu }(q)\ll \Lambda _{m}^{\mathrm{II}}(q),\ \ \ \ \ \ |\Lambda
_{2}^{\mu \mathrm{I}}(q)|^{2}=|\Lambda _{2}^{\mathrm{I}\mu }(q)|^{2}\ll
\Lambda _{2}^{\mathrm{II}}(q),
\end{equation}
where $m=1,2.$

\subsubsection*{Quantum Fokker-Planck master equation}

Similarly to the previous analysis [second-third case, Eq.~(\ref{ThirdSecond}
)] the approximation~(\ref{FourthApproxy}) does not guarantees the
positivity on the state $\varrho _{t}(q)$ at short time and small length
scales [here defined from $\lambda ^{\mathrm{II}}(r|q),$ see conditions~(\ref
{ShortScales}) and model~(\ref{ExampleRate})]. The lack of positivity
originates from \ quantum-classical \textit{coupling terms} that represent
\textquotedblleft finite\textquotedblright\ discrete (underlying) changes in
the hybrid state $\varrho _{t}(q).$ These changes are induced by the
operators $(a_{\mu }\mathrm{I}_{s}+b_{\mu }V_{\mu })$ [see Eq.~(\ref
{FourthOperators})]. Consequently, the transformations associated to these
operators are \textit{infinitesimal}, approach the identity, if one assumes 
\begin{equation}
|b_{\mu }|\ll |a_{\mu }|\approx 1.  \label{SmallQuantum}
\end{equation}
Introducing these conditions in the definitions of the jump moments [Eqs.~( 
\ref{LambdaMom_1}) and~(\ref{LambdaMom_2})] expressed in terms of the
underlying rates [Eqs.~(\ref{Lamdba_1}) and~(\ref{Lamdba_2})], from Eq.~(\ref
{FourthApproxy}) we write 
\begin{eqnarray}
\frac{\partial \varrho _{t}(q)}{\partial t} &=&\Lambda _{0}^{\mu }(q)(V_{\mu
}\varrho _{t}(q)V_{\mu }^{\dagger }-\frac{1}{2}\{V_{\mu }^{\dagger }V_{\mu
},\varrho _{t}(q)\}_{+})  \notag \\
&&-\frac{\partial }{\partial q}[\Lambda _{1}^{\mathrm{II}}(q)\varrho
_{t}(q)]+\frac{1}{2}\frac{\partial ^{2}}{\partial q^{2}}[\Lambda _{2}^{
\mathrm{II}}(q)\varrho _{t}(q)],  \notag \\
&&-i[H_{0}(q),\varrho _{t}(q)]_{-}  \notag \\
&&-\frac{\partial }{\partial q}V_{\mu }(q)\varrho _{t}(q)-\frac{\partial }{
\partial q}\varrho _{t}(q)V_{\mu }^{\dagger }(q).  \label{QuantumFP}
\end{eqnarray}
Here, the Hamiltonian is defined by Eq.~(\ref{Hamiltonian_q}). For
simplifying the expression we introduced the operators $V_{\mu }(q)\equiv
\Lambda _{1}^{\mu \mathrm{I}}(q)V_{\mu }$ and $V_{\mu }^{\dagger }(q)\equiv
\Lambda _{1}^{\mathrm{I}\mu }(q)V_{\mu }^{\dagger }.$

The quantum Fokker-Planck master equation~(\ref{QuantumFP}) has fewer terms
than the diffusive approximation~(\ref{FourthApproxy}). This feature is a
consequence of maintaining only contributions up to second order in $|b_{\mu
}|$ \textit{and} the small coordinate scale $\Delta q\equiv \sqrt{\Lambda
_{2}^{\mathrm{II}}/(2\Lambda _{0}^{\mathrm{II}})}$ [see example~(\ref
{ExampleRate})]. The moment $\Lambda _{0}^{\mathrm{II}}$ defines the
minor-time scale of the dynamics~(\ref{FourthApproxy}). The moments $
|\Lambda _{0}^{\mathrm{I}\mu }|$ and $\Lambda _{0}^{\mu },$ with respect to $
\Lambda _{0}^{\mathrm{II}},$ are of first and second order in $|b_{\mu }|$
respectively. They define the Hamiltonian and Lindblad contributions in\
both expressions. The moment $\Lambda _{1}^{\mathrm{II}}$ (a
\textquotedblleft velocity\textquotedblright ) is proportional to $\Delta q.$
The moments $|\Lambda _{1}^{\mathrm{I}\mu }|$ and $\Lambda _{1}^{\mu }$ are
proportional to $|b_{\mu }|$\ and $|b_{\mu }|^{2}$ respectively. Thus, to
maintain second order contributions, the derivative proportional to $\Lambda
_{1}^{\mu }$ is disregarded in Eq.~(\ref{QuantumFP}). Finally, the moment $
\Lambda _{2}^{\mathrm{II}}$ (a diffusion coefficient) is proportional to $
\Delta q^{2}.$ Consequently, the contributions proportional to$\ |\Lambda
_{2}^{\mathrm{I}\mu }|$ and $\Lambda _{2}^{\mu }$ (order $|b_{\mu }|$\ and $
|b_{\mu }|^{2})$ are also disregarded. Hence, Eq.~(\ref{QuantumFP}) applies
when the underlying transformations associated to the quantum and classical
subsystems are infinitesimal,\ this property being scaled by $b_{\mu }$ and $
\Delta q.$

In order to guaranty positivity of the hybrid state, in Eq.~(\ref{QuantumFP}
) it must be fulfilled $\forall \mu $ that 
\begin{equation}
\Lambda _{0}^{\mu }(q)\neq 0,\ \ \ \ \ \ \ \Lambda _{1}^{\mathrm{I}\mu }(q) 
\frac{1}{\Lambda _{0}^{\mu }(q)}\Lambda _{1}^{\mu \mathrm{I}}(q)\leq \Lambda
_{2}^{\mathrm{II}}(q).  \label{InequalityContinuo}
\end{equation}
This inequality is a direct consequence of the completely positive condition
defined by Eq.~(\ref{diagonal}). It expresses that the \textquotedblleft
quadratic velocity\textquotedblright\ associated to the diagonal
contribution, $\Lambda _{2}^{\mathrm{II}}(q)\Lambda _{0}^{\mu }(q),$ must be
larger than the \textquotedblleft quadratic velocity\textquotedblright\
associated to the \textquotedblleft non-diagonal\textquotedblright\ coupling
mechanisms, $|\Lambda _{1}^{\mu \mathrm{I}}(q)|^{2}=|\Lambda _{1}^{\mathrm{I}
\mu }(q)|^{2}.$ Notice that the fourth case [Eq.~(\ref{FourthOperators})]
corresponds to $|\Lambda _{1}^{\mathrm{I}\mu }(q)|^{2}=\Lambda _{0}^{\mu
}(q)\Lambda _{2}^{\mathrm{II}}(q).$ Thus, the inequality in Eq.~(\ref
{InequalityContinuo}) takes into account the possibility of the joint action
of the second, third and fourth coupling mechanisms [Eq.~(\ref
{DosTresDifusiva})]. In this situation Eq.~(\ref{FourthApproxy}) can also be
applied to the general case defined by Eq.~(\ref{Matrizon}).

We remark that the structure~(\ref{QuantumFP}) and the condition~(\ref
{InequalityContinuo}) are consistent with the results of Refs.~\cite
{oppen1,oppen2,lastDiosi,QFP}. The general non-diagonal case, index $\mu
\rightarrow (\mu ,\nu ),$ can be recovered straightforwardly from these
expressions.

\section{Examples}

Here we study a set of dynamics that corroborate our main results. The
hybrid state [Eq.~(\ref{Suma})] has the structure $\Xi _{t}=\sum_{n}\rho
_{n}(t)\otimes |n\rangle \langle n|.$ The basis $\{|n\rangle \}_{-\infty
}^{+\infty }$ corresponds to the classical subsystem while the quantum
subsystem is a two-level system, being characterized by the conditional
(unnormalized) states 
\begin{equation}
\rho _{n}(t)=\left( 
\begin{array}{cc}
p_{n}^{+}(t) & c_{n}(t) \\ 
c_{n}^{\ast }(t) & p_{n}^{-}(t)
\end{array}
\right) .  \label{MatrixElements}
\end{equation}
These matrix elements are defined in a basis $\{|\pm \rangle \}$ of the
two-level system. The initial condition reads 
\begin{equation}
\rho _{n}(0)=\left( 
\begin{array}{cc}
\langle +|\rho _{0}|+\rangle & \langle +|\rho _{0}|-\rangle \\ 
\langle -|\rho _{0}|+\rangle & \langle -|\rho _{0}|-\rangle
\end{array}
\right) p_{n}(0),
\end{equation}
where $\rho _{0}$ and $\{p_{n}(0)\}_{-\infty }^{+\infty }$ are the initial
conditions of the quantum and classical subsystems respectively. For
simplicity we assume $p_{n}(0)=\delta _{n,n_{0}},$ where $n_{0}$ is an
arbitrary position in the discrete (classical) space.

The examples studied below are defined in terms of a set of quantum and
classical superoperators. In the quantum Hilbert space we denote a Lindblad
contribution as 
\begin{equation}
\mathcal{L}_{V}[\rho ]\equiv V\rho V^{\dag }-\frac{1}{2}\{V^{\dag }V,\rho
\}_{+},  \label{LindbladDef}
\end{equation}
which depends on the operator $V$ while $\rho $ is an arbitrary matrix in
the quantum Hilbert space. In the classical space, we introduce a discrete
\textquotedblleft force\textquotedblright\ operator 
\begin{equation}
\mathrm{L}_{1}[f_{n}]\equiv \frac{1}{2}(f_{n+1}-f_{n-1}),
\label{L1_Discreto}
\end{equation}
while a discrete \textquotedblleft diffusion\textquotedblright\ operator is
defined as 
\begin{equation}
\mathrm{L}_{2}[f_{n}]\equiv f_{n+1}+f_{n-1}-2f_{n}.  \label{L2_Discreto}
\end{equation}
Both operators act on a space of countable functions $\{f_{n}\}_{-\infty
}^{+\infty }.$ We notice that the quantum and classical operators commutate,
that is, $\mathcal{L}_{A}\mathrm{L}_{1}=\mathrm{L}_{1}\mathcal{L}_{A}$ and $
\mathcal{L}_{A}\mathrm{L}_{2}=\mathrm{L}_{2}\mathcal{L}_{A}.$

A \textit{continuous limit} is defined by introducing the coordinate $
q\equiv nr_{0},$ where $r_{0}$ is the small length scale of the problem. The
hybrid state [Eq.~(\ref{HybridStateContinuo})] becomes $\Xi _{t}=\int
dq\varrho _{q}(t)\otimes |q\rangle \langle q|,$ where the (density of)
quantum state, $\varrho _{q}(t)dq=r_{0}\rho _{n}(t),$ is denoted as 
\begin{equation}
\varrho _{q}(t)=\left( 
\begin{array}{cc}
P_{q}^{+}(t) & C_{q}(t) \\ 
C_{q}^{\ast }(t) & P_{q}^{-}(t)
\end{array}
\right) .  \label{DensityTLS}
\end{equation}
In the continuos limit, the operators~(\ref{L1_Discreto}) and~(\ref
{L2_Discreto}) are replaced by 
\begin{equation}
\mathrm{L}_{1}\rightarrow r_{0}\frac{\partial }{\partial q},\ \ \ \ \ \ \ 
\mathrm{L}_{2}\rightarrow r_{0}^{2}\frac{\partial ^{2}}{\partial ^{2}q},
\label{OperatorAprox}
\end{equation}
when the a diffusive limit is approached.

\subsection{Dephasing correlated random walk}

In the first example we consider the time evolution 
\begin{eqnarray}
\frac{d\rho _{n}}{dt} &=&\gamma (\sigma _{z}\rho _{n+1}\sigma _{z}+\sigma
_{z}\rho _{n-1}\sigma _{z}-2\rho _{n})  \notag \\
&&+\phi (\rho _{n+1}+\rho _{n-1}-2\rho _{n}).  \label{DephasingCorrelated}
\end{eqnarray}
For simplicity, from now on we omit the time-dependence of all statistical
objects; here $\rho _{n}(t)\rightarrow \rho _{n}.$ With $\sigma _{z}$ we
denote the $z$-Pauli matrix in the basis $\{|\pm \rangle \}$ of the
two-level system.

It is simple to realize that the dynamics~(\ref{DephasingCorrelated})
correspond to a combined action of the second [Eq.~(\ref{Dos})] and third
coupling mechanisms [Eq.~(\ref{Tres})] with rates $\phi $ and $\gamma $
respectively. Notice that in the third mechanism any classical transition $
n^{\prime }=n\pm 1\rightarrow n$ is endowed by the quantum transformation $
\rho _{n}\rightarrow \sigma _{z}\rho _{n^{\prime }}\sigma _{z}.$ The
superoperator $(\sigma _{z}\bullet \sigma _{z})$ changes the sign of the
coherences of $\rho _{n\pm 1}$ while maintains invariant its populations.
Thus, \textit{dephasing is inherently correlated with the jumps in the
classical space}.

The hybrid evolution~(\ref{DephasingCorrelated}) can straightforwardly be
rewritten as 
\begin{equation}
\frac{d\rho _{n}}{dt}=2\gamma \mathcal{L}_{\sigma _{z}}[\rho _{n}]+\phi 
\mathrm{L}_{2}[\rho _{n}]+\gamma \mathrm{L}_{2}[\sigma _{z}\rho _{n}\sigma
_{z}],  \label{PrimerExampleDiscreto}
\end{equation}
where the superoperators are defined by Eqs.~(\ref{LindbladDef}) and~(\ref
{L2_Discreto}). Written in terms of the matrix elements~(\ref{MatrixElements}
), it follows that the populations evolves as 
\begin{equation}
\frac{dp_{n}^{(s)}}{dt}=(\phi +\gamma )\mathrm{L}_{2}[p_{n}^{(s)}],\ \ \ \ \
\ \ s=\pm 1,  \label{PoblacionDiscreta}
\end{equation}
while for the coherences we get 
\begin{equation}
\frac{dc_{n}}{dt}=-4\gamma c_{n}+(\phi -\gamma )\mathrm{L}_{2}[c_{n}].
\label{CoherenciaDiscreta}
\end{equation}

The above time-evolutions can be solved in an exact way by means of a
characteristic function approach~\cite{characteristic}. For the populations,
from Eq.~(\ref{PoblacionDiscreta}) we get the solution 
\begin{equation}
p_{n}^{(s)}(t)=e^{-2(\phi +\gamma )t}\ \mathrm{I}_{|n-n_{0}|}[2(\phi +\gamma
)t]\ \langle s|\rho _{0}|s\rangle ,  \label{PopusSolDiscreto}
\end{equation}
where $s=\pm 1,$ $n_{0}$ labels the initial position of the classical
subsystem, and $\mathrm{I}_{n}(x)$ is the nth modified Bessel function of
the first kind. We notice that this expression correspond to the solution of
a classical random walk on a discrete space~\cite{kampen} with
characteristic rate $(\phi +\gamma ).$

For the coherences, from Eq.~(\ref{CoherenciaDiscreta}) it follows that 
\begin{equation}
c_{n}(t)=e^{-4\gamma t}\{e^{-2(\phi -\gamma )t}\ \mathrm{I}
_{|n-n_{0}|}[2(\phi -\gamma )t]\}\ \langle +|\rho _{0}|-\rangle .
\label{CoherSolDiscreto}
\end{equation}
While this solution has the same structure than Eq.~(\ref{PopusSolDiscreto}
), the argument of the Bessel function depends on $(\phi -\gamma )\gtreqless
0.$ There is not any constraint on this difference of underlying rates.
Using that $\mathrm{I}_{n}(-x)=(-1)^{n}\mathrm{I}_{n}(x),$ we notice that
the coherences develops a change of sign between neighboring sites when the
third mechanism dominates on the second one, $\phi <\gamma .$ In spite of
this property, the conditional states $\{\rho _{n}(t)\}$ [Eq.~(\ref
{MatrixElements})] are positive definite $\forall n.$ In fact, given that $
p_{n}^{(s)}(t)>0$\ $(s=\pm 1),$ the positivity of $\rho _{n}(t)$ is
fulfilled when its determinant, denoted as $\det [\rho _{n}(t)],$ is a
positive quantity, 
\begin{equation}
\det [\rho _{n}(t)]=p_{n}^{(+)}(t)p_{n}^{(-)}(t)-|c_{n}(t)|^{2}\geq 0.
\label{DeterminanteDiscreto}
\end{equation}
This condition is always fulfilled when taking the matrix elements defined
by the solutions~(\ref{PopusSolDiscreto})\ and~(\ref{CoherSolDiscreto}).

\subsubsection*{Diffusive approximation}

Here we study a diffusive approximation to the hybrid master equation~(\ref
{DephasingCorrelated}). Taking into account the notation~(\ref{DensityTLS}),
from the equivalent master equation~(\ref{PrimerExampleDiscreto}), we write
the approximation $[\varrho _{q}(t)\rightarrow \varrho _{q}]$ 
\begin{equation}
\frac{\partial \varrho _{q}}{\partial t}\approx 2\gamma \mathcal{L}_{\sigma
_{z}}[\varrho _{q}]+\frac{D_{\phi }}{2}\frac{\partial ^{2}}{\partial ^{2}q}
\varrho _{q}+\frac{D_{\gamma }}{2}\frac{\partial ^{2}}{\partial ^{2}q}\Big{(}
\sigma _{z}\varrho _{q}\sigma _{z}\Big{)}.  \label{DiffusiveApproxFirstExam}
\end{equation}
Here, the diffusion coefficients, from the approximation~(\ref{OperatorAprox}
), read 
\begin{equation}
D_{\phi }=2\phi r_{0}^{2},\ \ \ \ \ \ \ \ D_{\gamma }=2\gamma r_{0}^{2}.
\label{DiffusionCoeficient23}
\end{equation}
The time-evolution~(\ref{DiffusiveApproxFirstExam}) has the structure
obtained when analyzing the combined action of the second and third coupling
mechanisms, Eq.~(\ref{ThirdSecond}).

The explicit solution of Eq.~(\ref{DiffusiveApproxFirstExam}) can be written
in terms of a Gaussian propagator, 
\begin{equation}
\mathbb{G}_{D}[q,t|\sigma _{0}]\equiv \sqrt{\frac{1}{2\pi (\sigma
_{0}^{2}+Dt)}}\exp \Big{[}-\frac{(q-q_{0})^{2}}{2(\sigma _{0}^{2}+Dt)}\Big{]}
.  \label{GaussPropagator}
\end{equation}
Here, $q_{0}$ and $\sigma _{0}$ denote initial mean value and standard
deviation respectively, while $D$ is a diffusion coefficient. The propagator
fulfills the diffusion equation $(\partial /\partial t)\mathbb{G}
_{D}[q,t|\sigma _{0}]=(D/2)(\partial ^{2}/\partial ^{2}q)\mathbb{G}
_{D}[q,t|\sigma _{0}].$ The populations of $\varrho _{q}(t)$\ [Eq.~(\ref
{DensityTLS})] can be written as 
\begin{equation}
P_{q}^{(s)}(t)=\mathbb{G}_{D_{\phi }+D_{\gamma }}[q,t|\sigma _{0}]\ \langle
s|\rho _{0}|s\rangle ,\ \ \ \ s=\pm 1,  \label{PsolDif1}
\end{equation}
while the coherence reads 
\begin{equation}
C_{q}(t)=e^{-4\gamma t}\mathbb{G}_{D_{\phi }-D_{\gamma }}[q,t|\sigma _{0}]\
\langle +|\rho _{0}|-\rangle .  \label{CohersolDif1}
\end{equation}

The solutions for the populations correspond to a standard diffusive
approximation to the classical random walk defined by Eq.~(\ref
{PoblacionDiscreta}). They always satisfies $0\leq P_{q}^{(s)}(t)\leq 1$ $
(s=\pm 1)$ and the normalization $P_{q}^{(+)}(t)+P_{q}^{(-)}(t)=1.$ They
provide a valid approximation independently of the value of the underlying
characteristic parameters $\phi $ and $\gamma ,$ equivalently, $D_{\phi }$
and $D_{\gamma }.$ On the other hand, consistently with the general
constraints (\ref{ThirdConstraints}), the solution for the coherences
imposes the condition 
\begin{equation}
D_{\phi }>D_{\gamma }.  \label{ConditionDiffusion}
\end{equation}
In this regime, the coherences performs a standard diffusion process that is
very well approximated by Eq.~(\ref{CohersolDif1}). Nevertheless, when $
D_{\gamma }>D_{\phi }$ the solution~(\ref{CohersolDif1}) \textit{diverges in
time}. The origin of this unphysical behavior reflects the change of sign of
the discrete solution $\{c_{n}(t)\}$ in a scale of order $r_{0}$ [Eqs.~(\ref
{CoherenciaDiscreta}) and~(\ref{CoherSolDiscreto})]. This property cannot be
recovered in a diffusive approximation, confirming in addition that here the
third coupling mechanism $(\phi =0,\ \gamma >0)$ cannot be approximated in
terms of the first and second derivatives of the conditional state $\varrho
_{q}(t).$

When condition~(\ref{ConditionDiffusion}) is fulfilled, both the populations 
$P_{q}^{(s)}(t)$\ and coherence $C_{q}(t)$ are well behaved an provide a
very well approximation to the discrete objects in the regime $\phi t>1$
[see Eq.(\ref{ShortScales})]. Nevertheless, a physical solution is only
granted when $\varrho _{q}(t)$ is a positive definite matrix. This condition
is fulfilled when its determinant $\mathrm{Det}[\varrho _{q}(t)]$ is
positive, 
\begin{equation}
\mathrm{Det}[\varrho
_{q}(t)]=P_{q}^{(+)}(t)P_{q}^{(-)}(t)-|C_{q}(t)|^{2}\geq 0\ \ \ \ \forall q.
\label{DetContinuo}
\end{equation}
\begin{figure}[tb]
\centering
\includegraphics[height=7cm,bb=35 590 730 1155]{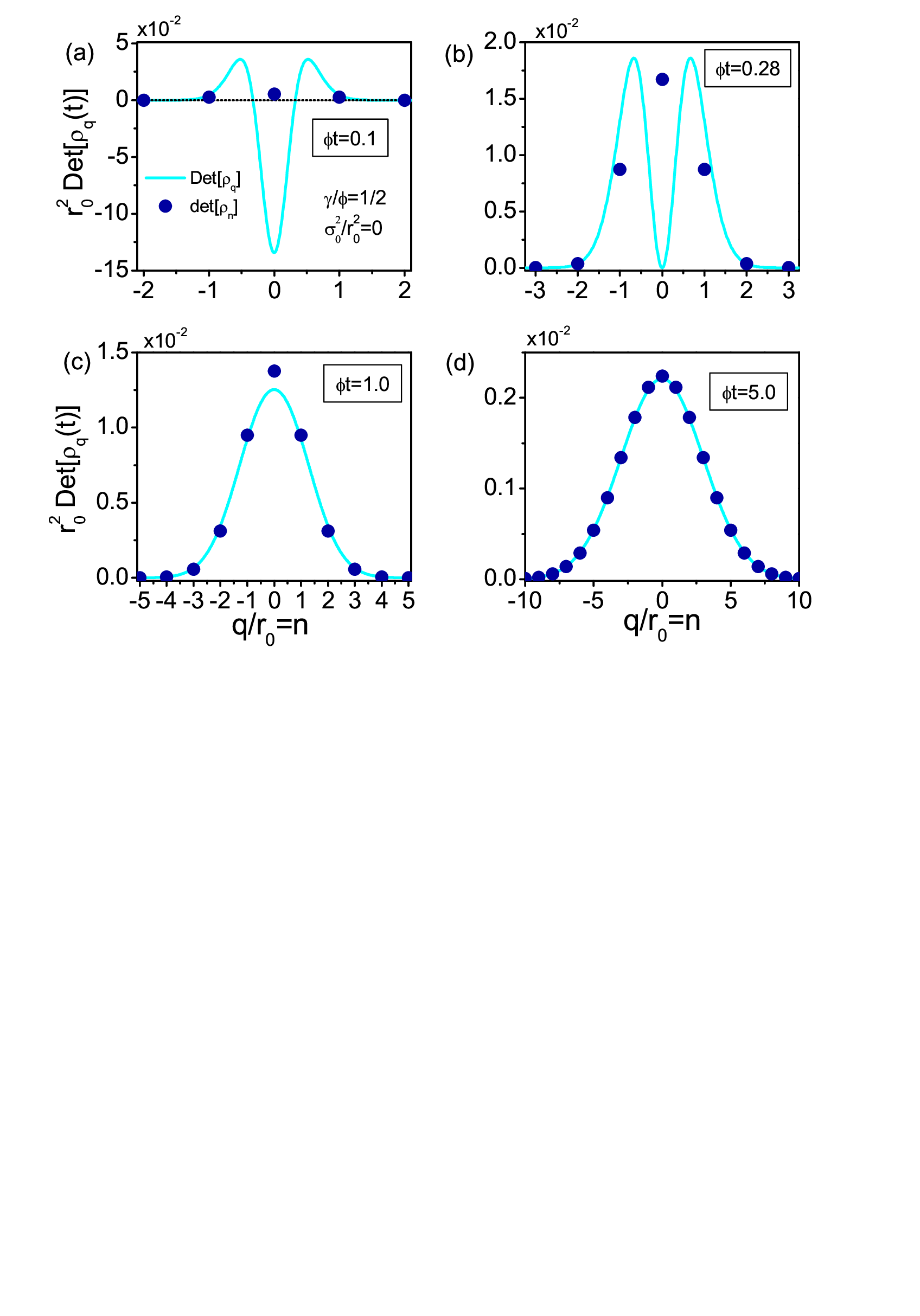}
\caption{Determinant of the conditional quantum state as a function of the
dimensionless coordinate $q/r_{0}=n$ at different times. The (blue) dots
correspond to the discrete dephasing random walk model Eq.~(\protect\ref
{DephasingCorrelated}) where $\det [\protect\rho _{n}(t)]$ is given by Eq.~(
\protect\ref{DeterminanteDiscreto}). The rates fulfills $\protect\gamma /
\protect\phi =2.$ The times are (a) $\protect\phi t=0.1,$ (b) $\protect\phi 
t=0.28,$ (c) $\protect\phi t=1.0,$ and (d) $\protect\phi t=5.0.$ The full
(cyan) lines correspond to the diffusive approximation, Eq.~(\protect\ref
{DiffusiveApproxFirstExam}), where $\mathrm{Det}[\protect\varrho _{q}(t)]$
is given by Eq.~(\protect\ref{DetContinuo}). The initial width is $\protect
\sigma _{0}^{2}/r_{0}^{2}=0,$ and $q_{0}/r_{0}=n_{0}=0.$}
\label{FiguraUno}
\end{figure}
In Fig.~\ref{FiguraUno} we plot both $\mathrm{Det}[\varrho _{q}(t)]$ and $
\det [\rho _{n}(t)]$ [Eq.~(\ref{DeterminanteDiscreto})] at different times.
The initial condition reads $\varrho _{q}(0)=\rho _{0}\delta (q),$ where the
quantum state $\rho _{0}$ is an eigenstate of the $x$-Pauli matrix, $\rho
_{0}=(1/2)\{\{1,1\},\{1,1\}\}.$ When $\phi t<1$ [Fig.~1(a)] at the origin $
(q=0)$ $\mathrm{Det}[\varrho _{q}(t)]$ develops a peak with negative
amplitude, which indicates an \textit{unphysical solution}. At later times
[Fig.~1(b)] at the origin $\mathrm{Det}[\varrho _{q}(t)]|_{q=0}=0,$ and
simultaneously it is positive $\forall q.$ For times $\phi t>1$
[Fig.~1(c)-(d)] $\mathrm{Det}[\varrho _{q}(t)]$ approaches the discrete
solution given by $\det [\rho _{n}(t)].$ In fact, when $\phi t\gg 1$ it is
expected that $\mathrm{Det}[\varrho _{nr_{0}}(t)]r_{0}^{2}\approx \det [\rho
_{n}(t)].$

The previous features demonstrate that the diffusive approximation~(\ref
{DiffusiveApproxFirstExam}) only provides physical solutions when $\phi t>1.$
This result is consistent with our previous general analysis [Eq.~(\ref
{ThirdConstraints})]. In order to characterize analytically this effect we
develops $\mathrm{Det}[\varrho _{q}(t)]$ around the initial condition $(q=0)$
and short times as 
\begin{equation}
(\mathrm{Det}[\varrho _{q}(t)])_{q=0}\approx \Big{(}\mathrm{Det}[\varrho
_{q}(0)]+t\Big{(}\frac{\partial }{\partial t}\mathrm{Det}[\varrho _{q}(t)]
\Big{)}_{t=0}\Big{)}_{q=0}.  \label{DetSeries}
\end{equation}
The most stringent condition emerges when the quantum subsystem starts in a
pure state. In fact, in this case the cero order vanishes identically, $\det
[\rho _{0}]=0\Rightarrow \mathrm{Det}[\varrho _{q}(0)]=0$ $\forall q.$ After
some algebra, demanding $(\mathrm{Det}[\varrho _{q}(t)])_{q=0}\geq 0$ it
follows the condition $t\geq t^{\ast }$ where 
\begin{equation}
\phi t^{\ast }=\frac{\phi }{8\gamma }\ln \Big{(}\frac{D_{\phi }+D_{\gamma }}{
D_{\phi }-D_{\gamma }}\Big{)}=\frac{\phi }{8\gamma }\ln \Big{(}\frac{\phi
+\gamma }{\phi -\gamma }\Big{)}<1.  \label{PositiveTime}
\end{equation}
Here, the second equality relies on the definitions~(\ref
{DiffusionCoeficient23}). This expression defines the time $t^{\ast }$ at
which the determinant at $q=0$ start to be positive. Furthermore, we
introduced the condition $\phi t^{\ast }<1,$\ which guarantees that
unphysical behaviors occur in the short time scale [notice that this
condition also guaranties the fulfillment of Eq.~(\ref{ConditionDiffusion}
)]. As shown in Fig.~\ref{FiguraUno}(b) when $\phi t\simeq \phi t^{\ast
}=0.274$ the determinant is positive not only at the origin but also for any
value of the classical coordinate. 
\begin{figure}[tb]
\centering
\includegraphics[height=7cm,bb=35 590 730 1155]{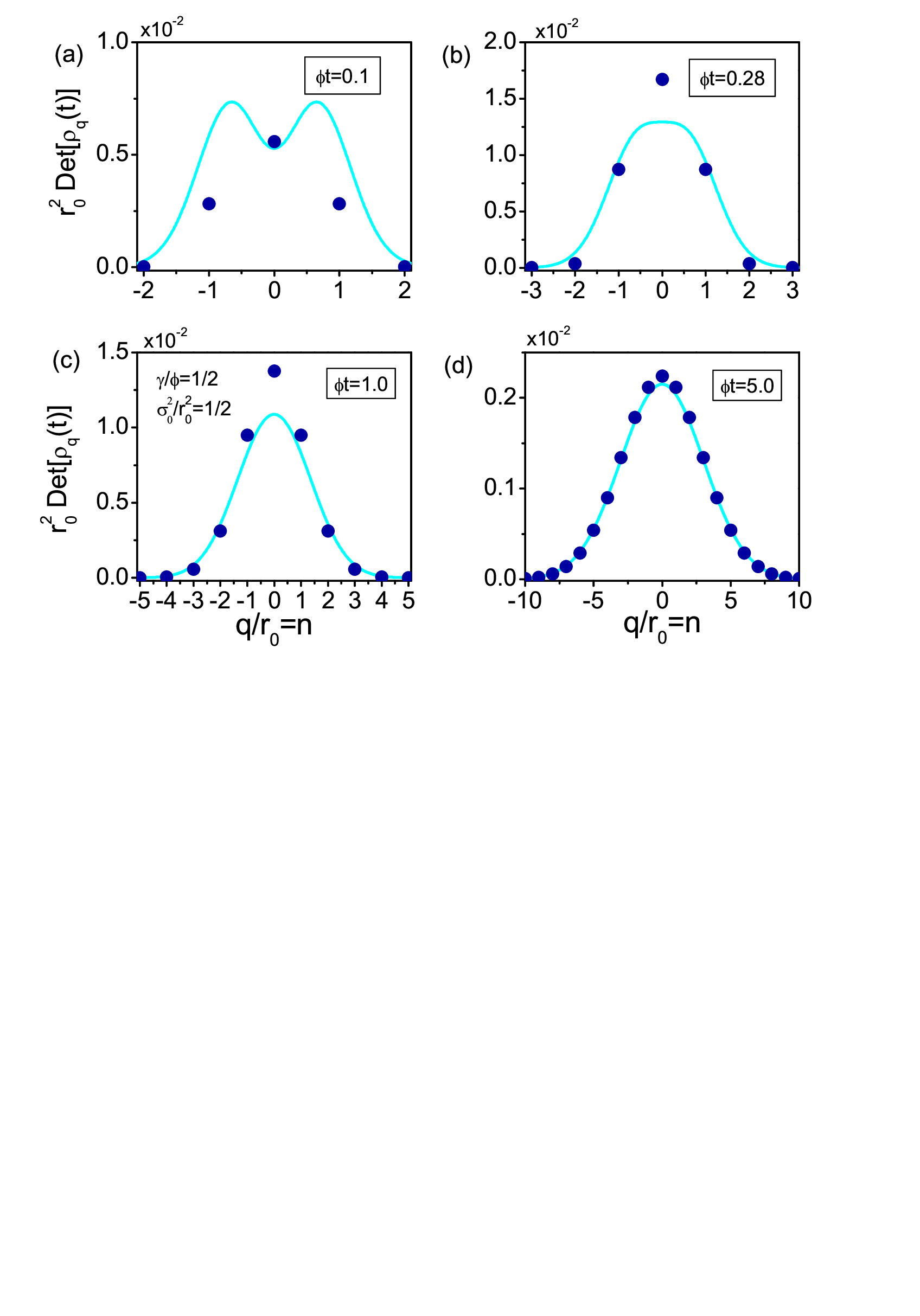}
\caption{Determinant of the conditional quantum state as a function of the
dimensionless coordinate $q/r_{0}=n$ at different times. The (blue) dots
correspond to the discrete dephasing random walk model Eq.~(\protect\ref
{DephasingCorrelated}) while the full (cyan) lines correspond to the
diffusive approximation Eq.~(\protect\ref{DiffusiveApproxFirstExam}). The
parameters are the same than in Fig.~1, but here the initial width of the
diffusive approximation is $\protect\sigma _{0}^{2}/r_{0}^{2}=1/2.$}
\label{FiguraDos}
\end{figure}

In the plots of Fig.~\ref{FiguraUno}, as well as in the derivation of~(\ref
{PositiveTime}) we assumed a totally localized initial condition $\varrho
_{q}(0)=\rho _{0}\delta (q).$ This assumption is consistent with the initial
condition of the underlying discrete model, $\rho _{n}(0)=\rho _{0}\delta
_{n,0}.$ Correspondingly, in the Gaussian propagator~(\ref{GaussPropagator})
we assumed a vanishing initial width, $\sigma _{0}=0.$ One can also assume a
finite initial value, $\sigma _{0}>0.$ By performing the same series
expansion [Eq.~(\ref{DetSeries})] than in the derivation of Eq.~(\ref
{PositiveTime}), we found that it is possible to \textit{guaranty the
positivity of }$Det[\varrho _{q}(t)]|_{0}$\textit{\ at short times if} 
\begin{equation}
\sigma _{0}^{2}\geq \frac{D_{\gamma }}{4\gamma }=\frac{r_{0}^{2}}{2}.
\label{AnchoPositividad}
\end{equation}
This result can be read as follows. The diffusive approximation becomes
physical on the short time scale $\phi t<1$ if instead of a localized
initial condition $[\sigma _{0}^{2}/r_{0}^{2}=0]$ a minimal initial
uncertainty is taken in the classical coordinate $[\sigma
_{0}^{2}/r_{0}^{2}=1/2].$ This feature has a clear physical meaning. In
fact, $D_{\gamma }/\gamma $ can be read as the classical (quadratic)
coordinate dispersion induced by the third coupling mechanism during its
characteristic time scale $1/\gamma .$ The isolated action of this mechanism
cannot be approximated by a diffusive dynamics. On the other hand, the
initial width $\sigma _{0}$ can be read as an initial spread induced by the
second coupling mechanism. In fact, the initial spread applies independently
of which direction in the quantum Hilbert space is taken into account. Thus,
the condition $\sigma _{0}^{2}\geq D_{\gamma }/(4\gamma )$ implies an
initial predominance of the second coupling mechanism over the third one,
which in turn leads to the positivity of $\mathrm{Det}[\varrho _{q}(t)]$ at
any time and space scales.

In Fig.~\ref{FiguraDos} we take the same model and parameter values than in
Fig.~\ref{FiguraUno} but taking an initial finite width $\sigma
_{0}^{2}/r_{0}^{2}=1/2.$ Consistently with the previous analysis, in the
short time scale $(\phi t<1)$ the determinant $\mathrm{Det}[\varrho _{q}(t)]$
is positive [Figs.~\ref{FiguraUno}(a)-(b)], while in the large time scale $
(\phi t>1)$ the influence of the initial width is negligible [Figs.~\ref
{FiguraUno}(c)-(d)]. In fact, these last plots are almost indistinguishable
with those of Fig.~\ref{FiguraUno}. Therefore, we conclude that the
diffusive approximation can lead to a positive definite solution when the
initial width of the classical subsystem is consistently restricted [Eq.~( 
\ref{AnchoPositividad})].

\subsection{Dephasing correlated random walk with non-diagonal coherent
couplings}

The second example can be read as a generalization of the previous one. The
dynamics is 
\begin{eqnarray}
\frac{d\rho _{n}}{dt} &=&\gamma (\sigma _{z}\rho _{n+1}\sigma _{z}+\sigma
_{z}\rho _{n-1}\sigma _{z}-2\rho _{n})  \notag \\
&&+\phi (\rho _{n+1}+\rho _{n-1}-2\rho _{n})  \notag \\
&&+\lambda ^{\mathrm{dn}}(\sigma _{z}\rho _{n+1}-\frac{1}{2}\{\sigma
_{z},\rho _{n}\}_{+})+h.c.  \notag \\
&&+\lambda ^{\mathrm{up}}(\sigma _{z}\rho _{n-1}-\frac{1}{2}\{\sigma
_{z},\rho _{n}\}_{+})+h.c.  \label{2ndExampleDiscreto}
\end{eqnarray}
In contrast to Eq.~(\ref{DephasingCorrelated}), here the third and fourth
lines take into account coupling mechanisms corresponding to the fourth case
[Eq.~(\ref{Cuatro})]. The general constraints that guaranty positivity [Eq.
( \ref{diagonal})] here read 
\begin{equation}
\frac{|\lambda ^{\mathrm{up}}|^{2}}{\gamma \phi }\leq 1,\ \ \ \ \ \ \ \ \ \ 
\frac{|\lambda ^{\mathrm{dn}}|^{2}}{\gamma \phi }\leq 1.
\label{ConstraintNODiagonal}
\end{equation}
Notice that each inequality defines an ellipse (strictly, a circle) in the
space defined by the complex parameters $\lambda ^{\mathrm{up}}$ and $
\lambda ^{\mathrm{dn}}.$

Using the general equivalent expression~(\ref{CuatroBis}), in terms of the
classical operators~(\ref{L1_Discreto}) and~(\ref{L2_Discreto}), the
evolution~(\ref{2ndExampleDiscreto}) can be rewritten as 
\begin{eqnarray}
\frac{d\rho _{n}}{dt}\! &=&\!(2\gamma \mathcal{L}_{\sigma _{z}}+\phi \mathrm{
\ L}_{2})[\rho _{n}]+\gamma \mathrm{L}_{2}[\sigma _{z}\rho _{n}\sigma
_{z}]-i \frac{\omega }{2}[\sigma _{z},\rho _{n}]_{-}  \notag \\
&&\!\!\!\!-(\delta \lambda \mathrm{L}_{1}[\sigma _{z}\rho _{n}]+h.c.)+(\frac{
\lambda }{2}\mathrm{L}_{2}[\sigma _{z}\rho _{n}]+h.c.).  \label{SecondL1L2}
\end{eqnarray}
The Lindblad superoperator $\mathcal{L}_{\sigma _{z}}$ is defined by Eq.~(
\ref{LindbladDef}). The parameters are 
\begin{equation}
\lambda \equiv \lambda ^{\mathrm{up}}+\lambda ^{\mathrm{dn}},\ \ \ \ \ \
\delta \lambda \equiv \lambda ^{\mathrm{up}}-\lambda ^{\mathrm{dn}},
\end{equation}
while the (real) frequency is 
\begin{equation}
\omega \equiv i(\lambda -\lambda ^{\ast }).
\end{equation}

From Eq.~(\ref{SecondL1L2}), the evolution of the quantum subsystem state, $
\rho _{t}=\sum \rho _{n},$ is 
\begin{equation}
\frac{d\rho }{dt}=2\gamma \mathcal{L}_{\sigma _{z}}[\rho ]-i\frac{\omega }{2}
[\sigma _{z},\rho ]_{-}.
\end{equation}
Thus, there is not any influence of the classical subsystem on the quantum
one [see Eq.~(\ref{QuantumIndependent})]. On the other hand, for writing the
evolution of the matrix elements of $\rho _{n}$ [Eq.~(\ref{MatrixElements}
)], in advance, we introduce the real and imaginary parts of the
characteristic parameters, 
\begin{equation}
\lambda =\lambda _{\mathrm{R}}+i\lambda _{\mathrm{I}},\ \ \ \ \ \ \ \ \
\delta \lambda =\delta \lambda _{\mathrm{R}}+i\delta \lambda _{\mathrm{I}}.
\end{equation}
For the populations we get 
\begin{equation}
\frac{dp_{n}^{(s)}}{dt}=-\alpha _{1}^{(s)}\mathrm{L}_{1}[p_{n}^{(s)}]+\alpha
_{2}^{(s)}\mathrm{L}_{2}[p_{n}^{(s)}],\ \ \ \ s=\pm 1,  \label{Pop2Discreta}
\end{equation}
where the coefficients are 
\begin{equation}
\alpha _{1}^{(s)}=2s\delta \lambda _{\mathrm{R}},\ \ \ \ \ \ \ \alpha
_{2}^{(s)}=\gamma +\phi +s\lambda _{\mathrm{R}}\geq 0.
\end{equation}
From the constraints~(\ref{ConstraintNODiagonal}) it follows the
inequalities $-4\sqrt{\gamma \phi }\leq \alpha _{1}^{(s)}\leq 4\sqrt{\gamma
\phi }$ and $0\leq \alpha _{2}^{(s)}\leq (\sqrt{\phi }+\sqrt{\gamma })^{2}.$
On the other hand, the coherence evolution reads 
\begin{equation}
\frac{dc_{n}}{dt}=-(i\omega +4\gamma )c_{n}-i\beta _{1}\mathrm{L}
_{1}[c_{n}]+\beta _{2}\mathrm{L}_{2}[c_{n}],  \label{Coher2Discreta}
\end{equation}
where the coefficients are 
\begin{equation}
\beta _{1}=2\delta \lambda _{\mathrm{I}},\ \ \ \ \ \ \ \ \ \beta _{2}=\phi
-\gamma +i\lambda _{\mathrm{I}}.
\end{equation}
From Eq.~(\ref{ConstraintNODiagonal}) it follows the inequalities $-4\sqrt{
\gamma \phi }\leq \omega \leq 4\sqrt{\gamma \phi },$ $-4\sqrt{\gamma \phi }
\leq \beta _{1}\leq 4\sqrt{\gamma \phi }$ and $|\beta _{2}|^{2}\leq (\phi
+\gamma )^{2}.$

Unlike the previous example, here in the evolution of the populations [Eq.~(
\ref{Pop2Discreta})], both the discrete-force\ as well as the
discrete-diffusion coefficients, $\alpha _{1}^{(s)}$ and $\alpha _{2}^{(s)}$
respectively, depend on the state ($s$) of the quantum subsystem. On the
other hand, in the evolution of the coherences [Eq.~(\ref{Coher2Discreta})],
the discrete-force and discrete-diffusion coefficients, $-i\beta _{1}$\ and $
\beta _{2},$ are complex parameters.

The evolutions~(\ref{Pop2Discreta}) and~(\ref{Coher2Discreta}) can be solved
analytically by using that a set of functions $\{f_{n}(t)\}_{-\infty
}^{+\infty }$ that satisfy 
\begin{equation}
\frac{df_{n}}{dt}=\alpha f_{n+1}+\beta f_{n-1}-(\alpha +\beta )f_{n},
\label{AsymetricalRW}
\end{equation}
with initial condition $f_{n}(0)=\delta _{n,n_{0}},$ from a characteristic
function approach~\cite{characteristic}, are given by 
\begin{equation}
f_{n}(t)=e^{-(\alpha +\beta )t}\ \mathrm{I}_{n-n_{0}}[2t\sqrt{\alpha \beta }
] \Big{(}\frac{\beta }{\alpha }\Big{)}^{\frac{n-n_{0}}{2}}.
\end{equation}
Notice that Eq.~(\ref{AsymetricalRW}) can be rewritten as 
\begin{equation}
\frac{df_{n}}{dt}=-(\beta -\alpha )\mathrm{L}_{1}[f_{n}]+\frac{\alpha +\beta 
}{2}\mathrm{L}_{2}[f_{n}],
\end{equation}
which has the structure of Eqs.~(\ref{Pop2Discreta}) and~(\ref
{Coher2Discreta}). 
\begin{figure}[tb]
\centering
\includegraphics[width=7 cm,bb=63 45 420 570]{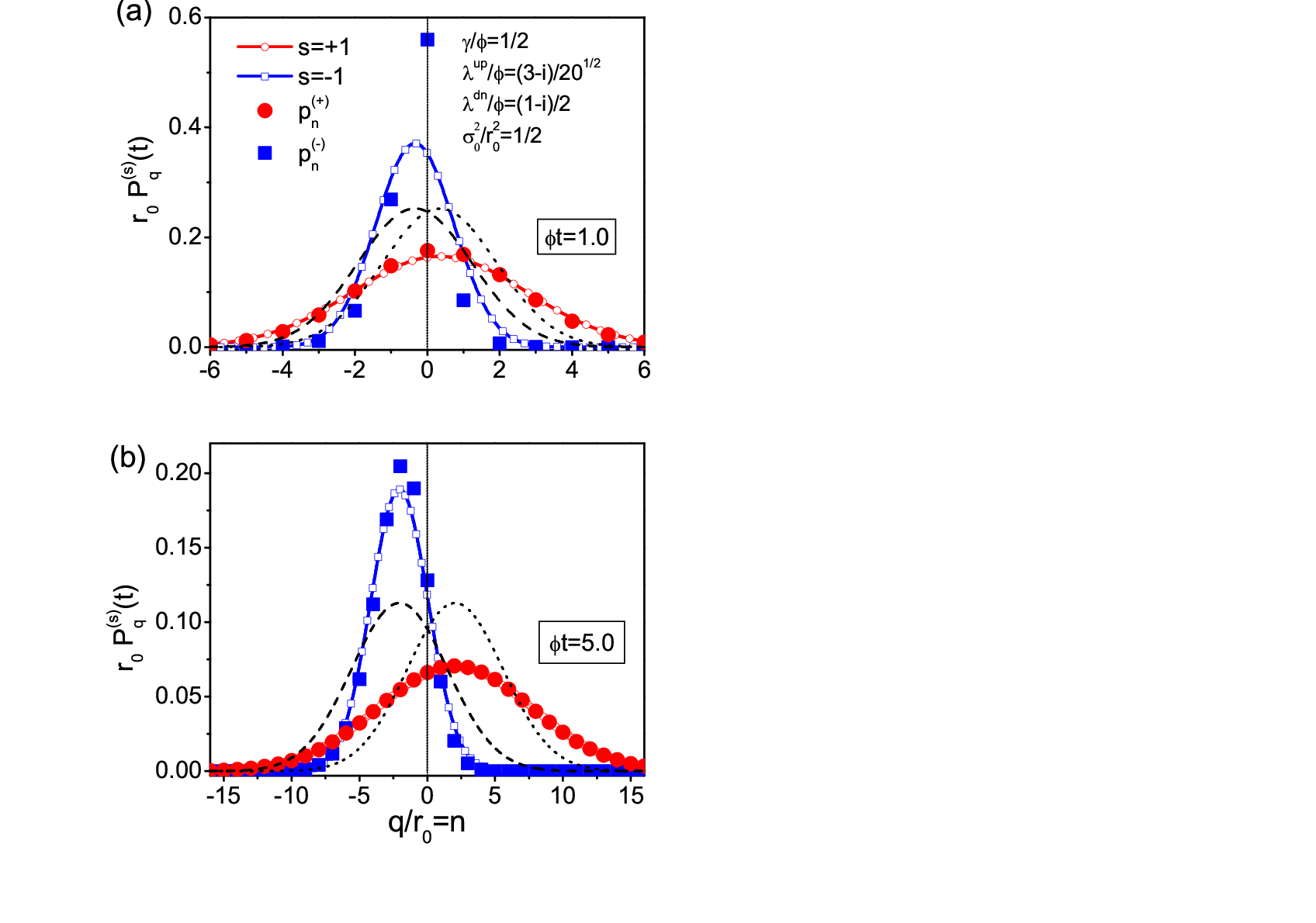}
\caption{Populations of the conditional quantum state as a function of the
dimensionless coordinate $q/r_{0}=n$ at different times. The (large) dots
correspond to the discrete dephasing random walk model Eq.~(\protect\ref
{2ndExampleDiscreto}) while the full lines correspond to the diffusive
approximation Eq.~(\protect\ref{DiffusiveApproxSecondExam}). The parameters
are $\protect\gamma /\protect\phi =1/2,$ $\protect\lambda ^{\mathrm{up}}/ 
\protect\phi =\protect\sqrt{1/20}(3-i),$ and $\protect\lambda ^{\mathrm{dn}
}/ \protect\phi =(1+i)/2.$ The initial width of the Gaussian approximation
is $\protect\sigma _{0}^{2}/r_{0}^{2}=1/2.$ The dotted and dashed lines
correspond to the solutions of the quantum Fokker-Planck dynamics, Eq.~( 
\protect\ref{QFP}).}
\label{FiguraTres}
\end{figure}

\subsubsection*{Diffusive approximation}

In a diffusive approximation, from Eq. (\ref{SecondL1L2}) we get 
\begin{eqnarray}
\frac{\partial \varrho _{q}}{\partial t} &\approx &2\gamma \mathcal{L}
_{\sigma _{z}}[\varrho _{q}]+\frac{D_{\phi }}{2}\frac{\partial ^{2}\varrho
_{q}}{\partial ^{2}q}+\frac{D_{\gamma }}{2}\frac{\partial ^{2}}{\partial
^{2}q}\Big{(}\sigma _{z}\varrho _{q}\sigma _{z}\Big{)}  \notag \\
&&-i\Big{(}\omega +F_{\lambda _{\mathrm{I}}}\frac{\partial }{\partial q}- 
\frac{D_{\lambda _{\mathrm{I}}}}{2}\frac{\partial ^{2}}{\partial ^{2}q} 
\Big{)}\frac{1}{2}[\sigma _{z},\varrho _{q}]_{-}  \notag \\
&&-\Big{(}F_{\lambda _{\mathrm{R}}}\frac{\partial }{\partial q}-\frac{
D_{\lambda _{\mathrm{R}}}}{2}\frac{\partial ^{2}}{\partial ^{2}q}\Big{)} 
\frac{1}{2}\{\sigma _{z},\varrho _{q}\}_{+},
\label{DiffusiveApproxSecondExam}
\end{eqnarray}
where as before [Eq.~(\ref{DiffusionCoeficient23})] the diffusion
coefficients are $D_{\phi }=2\phi r_{0}^{2},$\ and $D_{\gamma }=2\gamma
r_{0}^{2}.$ Furthermore, we introduced the extra coefficients associated to
the fourth coupling mechanism 
\begin{eqnarray}
F_{\lambda } &=&2\delta \lambda r_{0}=F_{\lambda _{\mathrm{R}}}+iF_{\lambda
_{\mathrm{I}}}, \\
D_{\lambda } &=&2\lambda r_{0}^{2}=D_{\lambda _{\mathrm{R}}}+iD_{\lambda _{ 
\mathrm{I}}}.
\end{eqnarray}
Written in terms of the matrix elements [Eq.~(\ref{DensityTLS})], for the
populations we get 
\begin{equation}
\frac{\partial P_{q}^{(s)}}{\partial t}\approx -sF_{\lambda _{\mathrm{R}}} 
\frac{\partial P_{q}^{(s)}}{\partial q}+\frac{D(s)}{2}\frac{\partial
^{2}P_{q}^{(s)}}{\partial ^{2}q},  \label{PopurContinuo2}
\end{equation}
where $D(s)=D_{\phi }+D_{\gamma }+sD_{\lambda _{\mathrm{R}}}.$ Similarly,
for the coherences we obtain 
\begin{equation}
\frac{\partial C_{q}}{\partial t}\approx -(i\omega +4\gamma
)C_{q}-iF_{\lambda _{\mathrm{I}}}\frac{\partial C_{q}}{\partial q}+\frac{ 
\tilde{D}}{2}\frac{\partial ^{2}C_{q}}{\partial ^{2}q},
\label{CoherolContinuo2}
\end{equation}
where $\tilde{D}=D_{\phi }-D_{\gamma }+iD_{\lambda _{\mathrm{I}}}.$ The
solution of Eqs.~(\ref{PopurContinuo2}) and~(\ref{CoherolContinuo2}) can be
written straightforwardly in terms of the Gaussian propagator~(\ref
{GaussPropagator}).

\begin{figure}[tb]
\centering
\includegraphics[width=7 cm,bb=63 45 420 570]{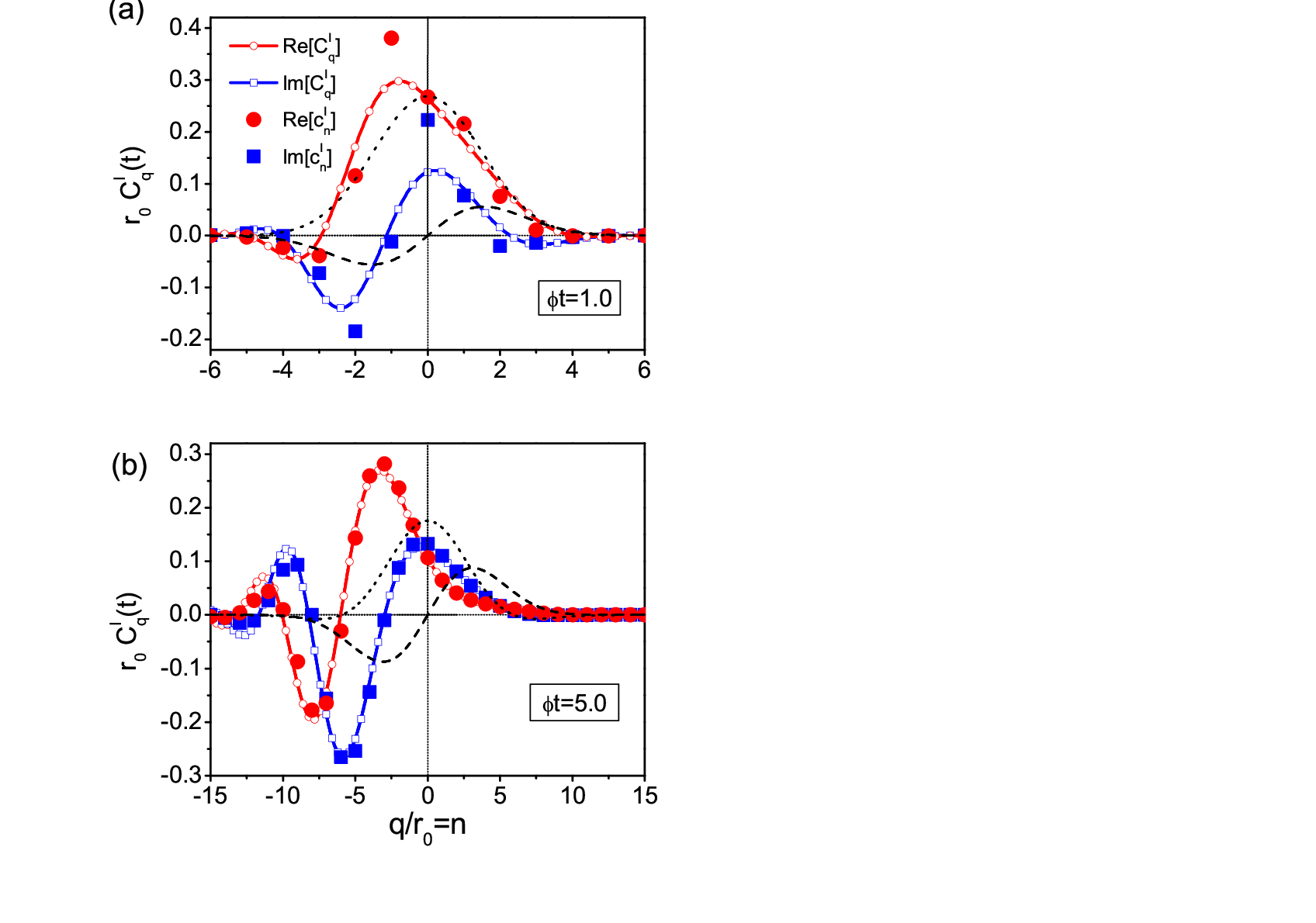}
\caption{Real and imaginary part of the coherence of the conditional quantum
state in an interaction representation (see text). The (large) dots
correspond to the discrete dephasing random walk model Eq.~(\protect\ref%
{2ndExampleDiscreto}) while the full lines correspond to the diffusive
approximation Eq.~(\protect\ref{DiffusiveApproxSecondExam}). The parameters
are the same than in Fig.~3. The dotted and dashed lines correspond to the
solutions of Eq.~(\protect\ref{QFP}).}
\label{FiguraCuatro}
\end{figure}

Taking the coordinate $q/r_{0}=n,$ in the time scale $\phi t\gg 1$ it is
expected that $\varrho _{nr_{0}}(t)r_{0}\approx \rho _{n}(t).$ In Fig.~\ref
{FiguraTres} we plot the population of the discrete (balls) and continuous
model (full lines), Eqs.~(\ref{Pop2Discreta}) and~(\ref{PopurContinuo2})
respectively. In Fig.~\ref{FiguraCuatro} we plot the real and imaginary
parts of the coherences corresponding to the discrete (balls) and continuous
model (full lines), Eqs.~(\ref{Coher2Discreta}) and~(\ref{CoherolContinuo2})
respectively. For clarity, the coherences are plotted in an interaction
representation, $c_{n}(t)=\exp [-(i\omega +4\gamma )t]c_{n}^{I}(t)$ and $
C_{q}(t)=\exp [-(i\omega +4\gamma )t]C_{q}^{I}(t).$ The initial conditions
read $\rho _{n}(0)=\mathbb{\delta }_{n,n_{0}}\rho _{0}$ and $\varrho
_{q}(0)= \mathbb{G}_{D}[q,0|\sigma _{0}]\rho _{0},$ where the quantum state
is again the pure state $\rho _{0}=(1/2)\{\{1,1\},\{1,1\}\}.$ The chosen
parameter values fulfill the equalities~(\ref{ConstraintNODiagonal}), that
is, $(|\lambda ^{\mathrm{up}}|/\phi )^{2}=(|\lambda ^{\mathrm{dn}}|/\phi
)^{2}=\gamma /\phi .$\ Hence, the dynamics only involves the fourth coupling
mechanism. 

From both figures we confirm that the diffusive approximation provides a
very good fitting to the discrete model even when $\phi t\approx 1.$ In
addition, Fig.~\ref{FiguraTres} explicitly shows that the state of the
quantum subsystem $(s=\pm 1)$ modify both the drift as well as the diffusion
coefficients of the conditional populations (center of and width of the
Gaussian densities respectively). Fig.~\ref{FiguraCuatro} explicitly shows
that even when the drift and diffusion coefficients are complex parameters,
the diffusive dynamics also provides a valid approximation to the underlying
conditional coherences. 
\begin{figure}[tb]
\centering
\includegraphics[height=7cm,bb=35 590 730 1155]{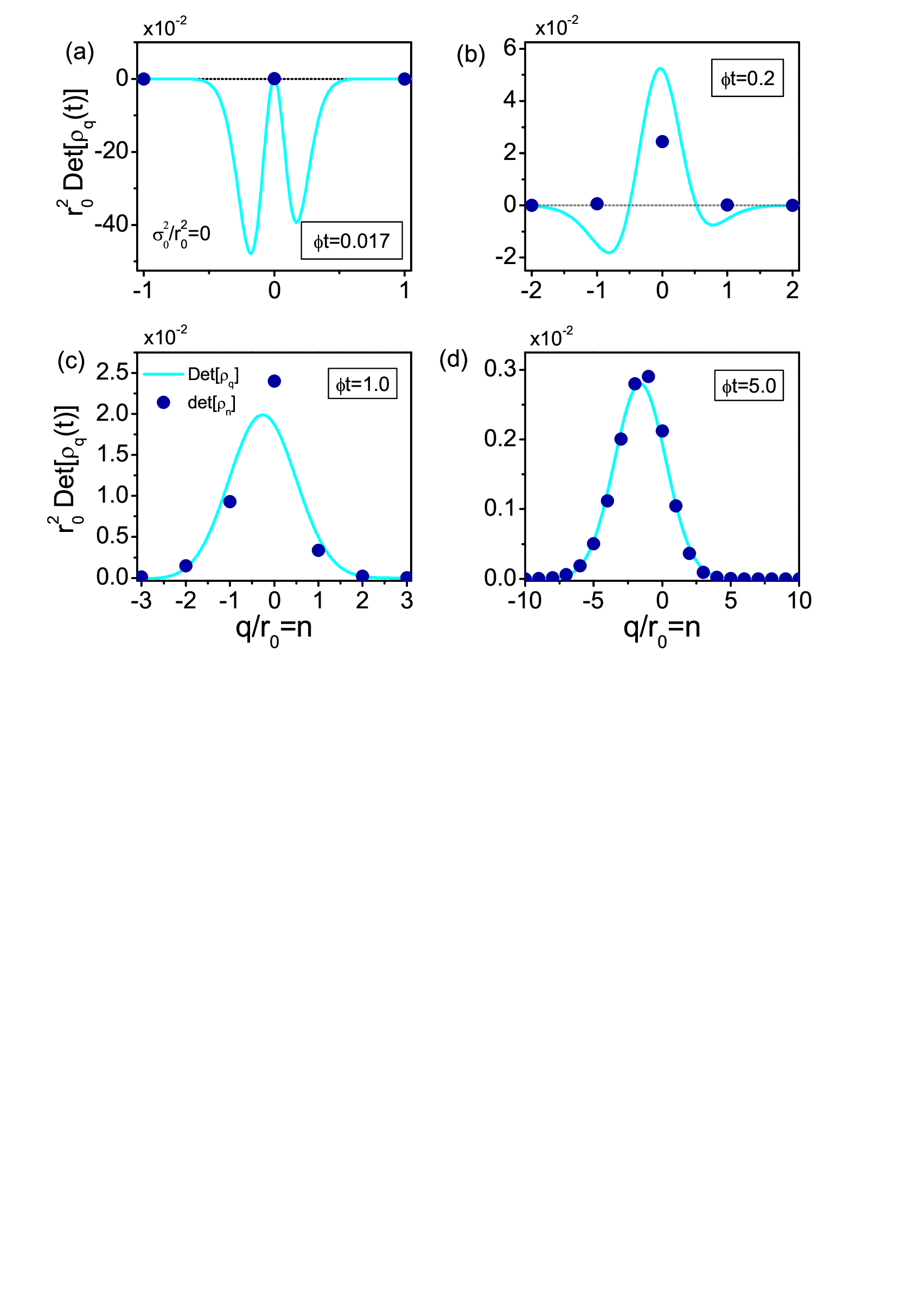}
\caption{Determinant of the conditional quantum state as a function of the
dimensionless coordinate $q/r_{0}=n$ at different times. The (blue) dots
correspond to the discrete dephasing random walk model Eq.~(\protect\ref
{2ndExampleDiscreto}) while the full (cyan) lines correspond to the
diffusive approximation Eq.~(\protect\ref{DiffusiveApproxSecondExam}). The
parameters are the same than in Fig.~3 and~4, but here the initial width of
the diffusive approximation is $\protect\sigma _{0}^{2}/r_{0}^{2}=0.$}
\label{FiguraCinco}
\end{figure}

In Figs.~\ref{FiguraTres} and~\ref{FiguraCuatro}, for the diffusive
approximation we taken and initial width $(\sigma _{0}/r_{0})^{2}=1/2.$ The
solutions with $(\sigma _{0}/r_{0})^{2}=0$ are almost indistinguishable in
the scale of the plots. Nevertheless, similarly to the previous example, the
initial width has a strong influence on the determinant of the conditional
states $\varrho _{q}(t).$ When $\sigma _{0}=0,$ the determinant $\mathrm{Det}
[\varrho _{q}(t)]$ [Eq.~(\ref{DetContinuo}) with the solutions of Eqs.~(\ref
{PopurContinuo2}) and~(\ref{CoherolContinuo2})] may assume negative
(unphysical) values. Taking $\sigma _{0}=0,$ developing $\mathrm{Det}
[\varrho _{q}(t)]|_{q=0}$ in a short time series expansion [Eq.~(\ref
{DetSeries})], and assuming a pure quantum initial condition $(\det [\rho
_{0}]=0,$ which implies $\mathrm{Det}[\varrho _{q}(0)]=0$ $\forall q)$ it
follows that the determinant at the origin is positive at times $t\geq
t^{\ast }$ where 
\begin{equation}
0\leq \phi t^{\ast }=\frac{\phi }{8\gamma }\frac{1}{(1-A)}\ln (B)<1.
\label{TimePlus2}
\end{equation}
Here, we introduced the dimensionless parameters 
\begin{equation}
A\equiv \frac{1}{8\gamma }\Big{[}\frac{F_{\lambda _{\mathrm{R}}}^{2}(D_{\phi
}+D_{\gamma })}{(D_{\phi }+D_{\gamma })^{2}-D_{\lambda _{\mathrm{R}}}^{2}}+ 
\frac{F_{\lambda _{\mathrm{I}}}^{2}(D_{\phi }-D_{\gamma })}{(D_{\phi
}-D_{\gamma })^{2}+D_{\lambda _{\mathrm{I}}}^{2}}\Big{]},
\end{equation}
and similarly 
\begin{equation}
B\equiv \sqrt{\frac{(D_{\phi }+D_{\gamma })^{2}-D_{\lambda _{\mathrm{R}
}}^{2} }{(D_{\phi }-D_{\gamma })^{2}+D_{\lambda _{\mathrm{I}}}^{2}}}.
\end{equation}

The time $t^{\ast }$ defined by Eq.~(\ref{TimePlus2}) can be read as a
generalization of Eq.~(\ref{PositiveTime}). The inequalities $0\leq \phi
t^{\ast }<1,$ \textit{which restrict the possible values of the
characteristic rates}, guaranty that unphysical behaviors associated to the
diffusive approximation [Eq.~(\ref{DiffusiveApproxSecondExam})] only occur
in the short time scale. On the other hand, if one assume an initial
non-vanishing width, it is possible to find out that the positivity of the
determinant at the origin is guaranteed at all times if $\sigma _{0}^{2}\geq
D_{\gamma }/4\gamma =r_{0}^{2}/2.$ This result is exactly the same than
Eq.~( \ref{AnchoPositividad}). Consequently, when the parameters are
consistent with $0<\phi t^{\ast }<1$ and taking $\sigma _{0}^{2}\geq
r_{0}^{2}/2$ we expect that the determinant will be positive at any time $t$
and any value of the classical coordinate $q.$ 
\begin{figure}[tb]
\centering
\includegraphics[height=7cm,bb=35 590 730 1155]{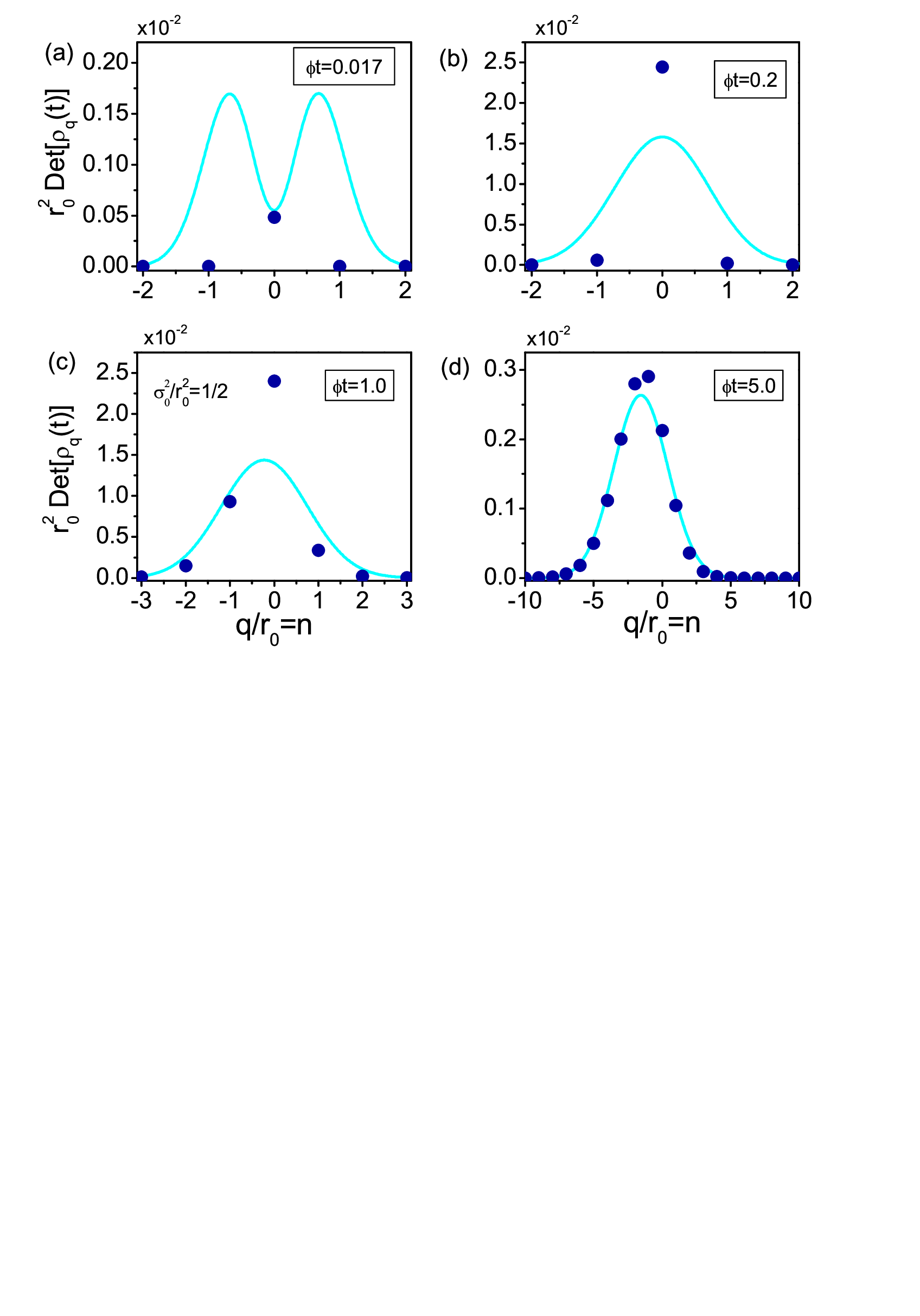}
\caption{Determinant of the conditional quantum state as a function of the
dimensionless coordinate $q/r_{0}=n$ at different times. The (blue) dots
correspond to the discrete dephasing random walk model Eq.~(\protect\ref
{2ndExampleDiscreto}) while the full (cyan) lines correspond to the
diffusive approximation Eq.~(\protect\ref{DiffusiveApproxSecondExam}). The
parameters are the same than in Fig.~3 and~4, where the initial width of the
diffusive approximation is $\protect\sigma _{0}^{2}/r_{0}^{2}=1/2.$}
\label{FiguraSeis}
\end{figure}

In order to check the previous conclusion in Fig.~\ref{FiguraCinco} we plot $
\mathrm{Det}[\varrho _{q}(t)]$ at different times taking a vanishing initial
width, $\sigma _{0}=0.$ Consistently when $\phi t\gtrsim \phi t^{\ast
}=0.0168$ the\ determinant at the origin becomes positive [Fig.~\ref
{FiguraCinco}(a)]. Nevertheless, in contrast with the previous example [see
Fig.~\ref{FiguraUno}(b)], due to the influence of the extra coupling
mechanism the initial negative value at $q=0$ has been \textquotedblleft
transported\textquotedblright\ to $|q|>0$ [Fig.~\ref{FiguraCinco}(a) and
(b)]. When $\phi t\gtrsim 1$ the determinant is positive for any value of $q$
[Fig.~\ref{FiguraCinco}(c) and (d)].

In Fig.~\ref{FiguraSeis} we plot $\mathrm{Det}[\varrho _{q}(t)]$ at
different times taking the initial width $\sigma _{0}^{2}=r_{0}^{2}/2.$
Consistently with the\ previous analysis we observe that now the determinant
is positive at any time and position. In addition, a good fitting of the
discrete model can be observed even in the transient regime $\phi t<1.$

\subsubsection*{Quantum Fokker-Planck master equation}

The lack of positivity of the determinant at the initial transient time
regime is avoided if one consider the quantum Fokker-Planck master dynamics
defined by Eq.~(\ref{QuantumFP}). For the present model, this limit can be
recovered from Eq.~(\ref{DiffusiveApproxSecondExam}) assuming valid the
conditions $D_{\phi }\gg D_{\gamma }$ and $D_{\phi }\gg |D_{\lambda }|,$
which under the replacement $D_{\gamma }\rightarrow 0$ and $D_{\lambda
}\rightarrow 0$ deliver 
\begin{eqnarray}
\frac{\partial \varrho _{q}}{\partial t} &=&2\gamma \mathcal{L}_{\sigma
_{z}}[\varrho _{q}]+\frac{D_{\phi }}{2}\frac{\partial ^{2}\varrho _{q}}{
\partial ^{2}q}-i\frac{\omega }{2}[\sigma _{z},\varrho _{q}]_{-}  \notag \\
&&-\Big{(}F_{\lambda }\frac{\partial }{\partial q}(\sigma _{z}\varrho
_{q})+h.c.\Big{)}.  \label{QFP}
\end{eqnarray}
The positivity constraint~(\ref{InequalityContinuo}) here lead to the
conditions $\gamma >0,$ $D_{\phi }>0,$ and $|F_{\lambda }|^{2}\leq 8\gamma
D_{\phi },$ which in turn are consistent with the underlying condition of
the discrete model, Eq.~(\ref{ConstraintNODiagonal}). Furthermore, they
imply $A\leq 1,$ $B=0,$ consequently $t^{\ast }=0$ [Eq.~(\ref{TimePlus2})].

In Figs.~\ref{FiguraTres} and~\ref{FiguraCuatro} we also plotted the
solution (dotted and dashed lines) corresponding to Eq.~(\ref{QFP}). The
evolution of the populations $P_{q}^{(s)}(t)$ and coherences $C_{q}(t)$\ can
be read straightforwardly from Eqs.~(\ref{PopurContinuo2}) and~(\ref
{CoherolContinuo2}) after taking $D_{\gamma }\rightarrow 0$ and $D_{\lambda
}\rightarrow 0.$ As clearly observed in Fig.~\ref{FiguraTres} the time
evolution~(\ref{QFP}) is unable to capture the dependence of the diffusion
coefficient with the state of the quantum system. In fact, both populations
have the same width. On the other hand, strong deviations can be observed in
Fig.~\ref{FiguraCuatro} when considering the real and imaginary part of the
coherences. In fact, Eq.~(\ref{QFP}), via de evolution~(\ref
{CoherolContinuo2}) with $D_{\gamma }\rightarrow 0$ and $D_{\lambda
}\rightarrow 0,$ implies the symmetries $\mathrm{Re}[C_{-q}(t)]=\mathrm{Re}
[C_{q}(t)]$ and $\mathrm{Im}[C_{-q}(t)]=-\mathrm{Im}[C_{q}(t)].$
Nevertheless, these properties are not fulfilled by the solution of the
underlying discrete model neither by its diffusive approximation based on
Eq.~(\ref{DiffusiveApproxSecondExam}).

\section{Summary and Conclusions}

We have introduced a deeper and general characterization of
quantum-classical hybrid dynamics. The study is based on embedding a
local-in-time Markovian hybrid dynamics in a bipartite Hilbert space where
the corresponding density matrix obeys a time-irreversible Lindblad equation
[Eq.~(\ref{Bipa})]. Requiring that one subsystem, at any time, must be
incoherent in a given fixed basis [Eq.~(\ref{Suma})] allowed us to recognize
the existence of four possible coupling mechanisms.

In the \textit{first coupling mechanism} [Eq.~(\ref{Uno})] the dynamics of
the quantum subsystem is given by a Lindblad dynamics whose parameters
depend on the state of the classical subsystem. There is not any backaction
on the classical subsystem, whose state is in fact a dynamical invariant
[Eq.~(\ref{NoBackActionClass_1})]. In the \textit{second coupling mechanism}
[Eq.~(\ref{Dos})] the classical subsystem obeys a classical master equation
[Eq.~(\ref{ClassicalMaster2}) ] while the state of the quantum subsystem
remains invariant [Eq.~(\ref{QuantumFrozen_2})]. Consequently, in this case
there is no backaction between the two subsystems. In the \textit{third
coupling mechanism} [Eq.~(\ref{Tres})] the transitions between the states of
each subsystem are correlated with the transitions of the other. Thus, both
subsystems influence each other. This symmetrical backaction precludes
writing a closed sub-dynamics. However, we also characterized the conditions
under which this is possible [Eqs.~(\ref{QuantumInd}) and~(\ref{popuInde})].
The \textit{fourth coupling mechanism} [Eq.~(\ref{Cuatro})]\ can be read as
a \textquotedblleft coherent superposition\textquotedblright\ of the second
and third cases. This property is derived from the structure of the
underlying bipartite coupling operators [Eq.~(\ref{FourthOperators})]. In
this case, there is always a backaction on the classical subsystem. Only
under specific conditions the quantum subsystem can evolves independently of
the previous one [Eq.~(\ref{QuantumIndependent})].

We \textit{concluded} that, in general, any quantum-classical hybrid
dynamics can be written in terms of the above four possible coupling
mechanisms. This result follows from the positivity of the underlying matrix
of rate coefficients [Eq.~(\ref{Matrizon})], property that in turn controls
the possible weights of the underlying coupling rates [Eq.~(\ref{diagonal})
and equivalently Eq.~(\ref{ConstraintPositivity})].

The above results allowed us to study in detail a \textit{diffusive
approximation}, where the evolution of the hybrid state is written only in
terms of the first and second derivatives with respect to a continuous
classical coordinate. Straightforwardly, the first mechanism can be extended
to the continuous case [Eq.~(\ref{UnoContinuo})]. Under standard conditions
on the transition rates, the second mechanism can always be modelled by a
diffusive mechanism [Eq.~(\ref{DosDifusiva})]. In fact, in this case the
coupling mechanism endow the classical subsystem with a time-irreversible
dynamics that is independent of the state of the quantum subsystem. Thus, a
diffusive process can be granted regardless of which direction in the
Hilbert space is considered. On the contrary, the third mechanism cannot be
addressed by a diffusive dynamics [Eq.~(\ref{PTresSeparada})]. In fact, the
underlying (coupled) quantum dynamics involves discrete changes which imply
that diffusion cannot be granted for arbitrary directions in Hilbert space.
Similarly to the fourth case, we conclude that a diffusive approximation
applies when the influence of the second mechanism is simultaneously
considered which is turn must be dominant over the others [Eqs.~(\ref
{ThirdSecond}) and~(\ref{FourthApproxy})].

When the underlying size of the (coupled) transformations undergone by the
quantum and classical subsystems can be taken as the small scale of the
problem, the hybrid evolution is given by a quantum Fokker-Planck master
equation [Eq.~(\ref{QuantumFP})]. These evolutions preserve at all times the
positivity of the hybrid state. However, they are unable to capture
dynamical effects that are well described in a diffusive approximation. In
this latter case the lack of positivity can only occurs in a short time
scale [Eqs.~(\ref{PositiveTime}) and~(\ref{TimePlus2})]. Moreover,
considering initial conditions for the classical system with a finite width
avoids the lack of positivity [Eq.~(\ref{AnchoPositividad})]. The previous
statements and conclusions were established from a set of representative
examples that admit and exact solution [Eqs.~(\ref{DephasingCorrelated})
and~(\ref{2ndExampleDiscreto})]. Dynamical effects such as the dependence of
diffusion coefficients on the state of the\ quantum subsystem as well as
complex diffusion coefficients are well described in a diffusive
approximation [Eqs.~(\ref{PopurContinuo2}) and~(\ref{CoherolContinuo2})]. In
this way, the present results establish the possibility of describing hybrid
dynamics with equations whose structure goes beyond quantum Fokker-Planck
master equations.

\section*{Acknowledgments}

The author thanks financial support from Consejo Nacional de 
Investigaciones Cient\'{\i}ficas y T\'{e}cnicas (CONICET), Argentina.

\appendix*

\section{Diffusive approximation of the fourth case\label{FourthCaseAppendix}}

In the continuous limit, the time-evolution corresponding to the fourth
coupling mechanism [Eq.~(\ref{Cuatro}) with $\lambda _{cc^{\prime }}^{\mu
\nu }=\delta _{\mu \nu }\lambda _{cc^{\prime }}^{\mu }],$ from the
equivalent expression~(\ref{CuatroBis}), can be written as 
\begin{eqnarray}
\frac{d\varrho _{t}(q)}{dt} &=&\Lambda _{0}^{\mu }(q)(V_{\mu }\varrho
_{t}(q)V_{\mu }^{\dagger }-\frac{1}{2}\{V_{\mu }^{\dagger }V_{\mu },\varrho
_{t}(q)\}_{+})  \notag \\
&&+(\digamma _{\mu }+\digamma _{I})[\varrho _{t}(q)]
\label{CuartoMasterSplit} \\
&&-i[H_{0}(q),\varrho _{t}(q)]+(\digamma _{\mu I}+\digamma _{I\mu })[\varrho
_{t}(q)].  \notag
\end{eqnarray}
The rates are defined in Eq.~(\ref{LambdaMom_1}) and~(\ref{LambdaMom_2}).
For shortening the expression we introduced the superoperator 
\begin{equation}
\digamma _{\mu }[\varrho _{t}(q)]\!=\!V_{\mu }\Big{\{}\int dr\lambda ^{\mu
}(r|q-r)\varrho _{t}(q-r)-\Lambda _{0}^{\mu }(q)\varrho _{t}(q)\Big{\}}
V_{\mu }^{\dagger }.
\end{equation}
Furthermore, 
\begin{equation}
\digamma _{I}[\varrho _{t}(q)]=\int dr\lambda ^{\mathrm{II}}(r|q-r)\varrho
_{t}(q-r)-\Lambda _{0}^{\mathrm{II}}(q)\varrho _{t}(q).
\end{equation}
The Hamiltonian is $H_{0}(q)=(i/2)\sum_{\mu }(\Lambda _{0}^{\mu \mathrm{I}
}(q)V_{\mu }-\Lambda _{0}^{\mathrm{I}\mu }(q)V_{\mu }^{\dagger }),$ while
the \textquotedblleft non-diagonal\textquotedblright\ contributions\ read 
\begin{equation}
\digamma _{\mu \mathrm{I}}[\varrho _{t}(q)]=\int dr\lambda ^{\mu \mathrm{I}
}(r|q-r)V_{\mu }\varrho _{t}(q-r)-\Lambda _{0}^{\mu \mathrm{I}}(q)V_{\mu
}\varrho _{t}(q),
\end{equation}
and similarly 
\begin{equation}
\digamma _{\mathrm{I}\mu }[\varrho _{t}(q)]=\int dr\lambda ^{\mathrm{I}\mu
}(r|q-r)\varrho _{t}(q-r)V_{\mu }^{\dagger }-\Lambda _{0}^{\mathrm{I}\mu
}(q)V_{\mu }^{\dagger }\varrho _{t}(q).
\end{equation}
For simplifying the expressions, an addition $\sum_{\mu }$ was not written
explicitly in the above equalities.

Assuming that the evolution~(\ref{CuartoMasterSplit}) is dominated by $
\digamma _{I}[\varrho _{t}(q)],$ this term as well as $\!\digamma _{\mu
}[\varrho _{t}(q)],$ $\digamma _{\mu \mathrm{I}}[\varrho _{t}(q)]$ and $
\digamma _{\mathrm{I}\mu }[\varrho _{t}(q)]=(\digamma _{\mu \mathrm{I}
}[\varrho _{t}(q)])^{\dagger }$ can be expanded in a second order Taylor
expansion [similar to Eq.~(\ref{Taylor})], which leads to Eq.~(\ref
{FourthApproxy}).


\begin{thebibliography}{99}
\bibitem{pie} For simplicity we considered that $\phi (r|q)=\phi (r).$ The
general case is covered by taking $\tau _{0}=\tau _{0}(q),$ $\delta \tau
_{0}=\delta \tau _{0}(q),$ and $r_{0}=r_{0}(q).$ On the other hand, the
general expressions for the moments are $\Phi _{m}(q)=m!r_{0}^{n}/\tau _{0}$
for $even$ $m\ $and $\Phi _{m}(q)=m!r_{0}^{n}/\delta \tau _{0}$ for $odd$ $
m. $
\bibitem{characteristic} Defining $F(\mathfrak{z},t)\equiv \sum_{n=-\infty
}^{n=+\infty }\mathfrak{z}^{n}f_{n}(t),$ the functions of interest $
\{f_{n}(t)\}_{-\infty }^{+\infty }$ read $f_{n}(t)=(2\pi i)^{-1}\oint 
\mathfrak{z}^{-n-1}F(\mathfrak{z},t),$ where the integral is taken around a
unit circle in the complex $\mathfrak{z}$-plane~\cite{kampen}.
\end{thebibliography}
\end{document}